\newif\ifOneCol

\ifOneCol
\documentclass[draftcls,12pt,onecolumn]{IEEEtran}
\else
\documentclass[journal]{IEEEtran}
\fi
\IEEEoverridecommandlockouts
\usepackage{mathtools}
\usepackage{graphicx}
\usepackage[noadjust]{cite}
\usepackage{amsmath}
\usepackage{textcomp}
\usepackage{tikz}
\usepackage[caption=false]{subfig}
\usepackage{array}
\usepackage{bm}
\usepackage{algorithm, algorithmicx}
\usepackage{algpseudocode}
\usepackage{mathrsfs}
\usepackage{amssymb}
\usepackage{enumerate}
\usepackage{ragged2e}
\usepackage{filecontents}
\usepackage{url}
\usepackage{times}
\usepackage{multirow}
\usepackage{soul}

\def\slpo{{\text{SL-PO}}}
\def\harq{{\text{HARQ}}}
\def\slinat{{\text{SL-INAT}}}
\def\sltx{{\text{SL-TX}}}
\def\slrx{{\text{SL-RX}}}
\def\sl{{\text{SL}}}
\def\nodata{{\text{NoData}}}
\def\txdata{{\text{TXData}}}
\def\rxdata{{\text{RXData}}}
\def\sam{{\text{SAM}}}
\def\samd{{\text{SAM-D}}}
\def\samu{{\text{SAM-U}}}

\def\e{{\mathrm{e}}}
\def\cdrx{{\text{CDRX}}}
\def\idrx{{\text{IDRX}}}
\def\cona{{\text{ConA}}}

\def\sltx{{\text{SL-TX}}}
\def\ue{{\text{UE}}}
\def\tx{{\text{TX}}}
\def\rx{{\text{RX}}}
\def\sldrx{{\text{SL-DRX}}}
\def\switch{{\text{switch}}}

\def\rrc{{\text{RRC}}}
\def\drxinat{{\text{DRX-INAT}}}
\def\datainat{{\text{Data-INAT}}}

\def\cdrx{{\text{CDRX}}}
\def\idrx{{\text{IDRX}}}

\def\data{{\text{Data}}}

\def\data{{\text{data}}}

\def\iat{{\text{IAT}}}

\ifCLASSINFOpdf

\else

\fi

\allowdisplaybreaks
\begin{document}
	
	\title{SCUBA: An In-Device Multiplexed Protocol for Sidelink Communication on Unlicensed Bands}
	
	\author{Vishnu Rajendran, Gautham Prasad, Lutz Lampe, \textit{Senior Member, IEEE}, Gus Vos, \textit{Senior Member, IEEE}
		\thanks{Vishnu Rajendran, Gautham Prasad, and Lutz Lampe are with the Department of Electrical and Computer Engineering, The University of British Columbia, Vancouver, BC, Canada. Gus Vos is with Sierra Wireless Inc., Richmond, BC, Canada. Email: vishnurc@ece.ubc.ca, gautham.prasad@alumni.ubc.ca, lampe@ece.ubc.ca, gvos@sierrawireless.com. }
		\thanks{A complementary precursor of this work is accepted for publication in the proceedings of the IEEE 92nd Vehicular Technology Conference: VTC2020-Fall, Victoria, BC, Canada~\cite{rajendran2020}.}
		\thanks{This work was supported by the Natural Sciences and Engineering Research Council of Canada (NSERC) and Sierra Wireless, Inc.}
	}

\markboth{Submission to IEEE IoT Journal}%
{Submitted paper}

\maketitle
	
\begin{abstract}
Device-to-device communication (D2D) is a key enabler for connecting devices together to form the Internet of Things (IoT). A growing issue with IoT networks is the increasing number of IoT devices congesting the spectral resources of the cellular bands. Operating D2D in unlicensed band alleviates this issue by offloading network traffic from the licensed bands, while also reducing the associated licensing costs. To this end, we present a new low-cost radio access technology (RAT) protocol, called Sidelink Communications on Unlicensed BAnds (SCUBA), which can be implemented on cellular devices such that it coexists with the legacy cellular protocol by operating as a secondary RAT in a time division multiplexed manner using the existing radio hardware. SCUBA is compatible on different types of cellular devices including the low-complexity half-duplex frequency division duplex machine type communication (MTC) user equipments. SCUBA provides flexible sidelink (SL) latency and battery life trade-off using a discontinuous reception procedure, which ensures that it is applicable across a wide range of use cases. We prove the effectiveness of our protocol with analyses and simulation results of the medium access control layer of SCUBA using different types of MTC traffic for both SL and the underlying cellular communication.

\end{abstract}
	
	
	\IEEEpeerreviewmaketitle
	
\section{INTRODUCTION}
\IEEEPARstart{D}{evice}-to-device (D2D) communication is a critical constituent of the Internet-of-Things (IoT). IoT and machine-type communication (MTC) devices interacting directly with each other on the sidelink (SL)\footnote{We use the terms SL communications and D2D communications interchangeably throughout the paper.}, in lieu of using uplink (UL) and downlink (DL) paths to talk via a central base station (BS), improves the communication latency and the device battery life~\cite{D2Dadv1,D2Dadv2,D2Dadv3,D2Dadv4}. D2D solutions are suitable in a wide range of MTC and IoT applications such as sensors, wearables, public lighting, container tracking, and smart metering. Additionally, using SL on unlicensed bands (SL-U) is a cost-effective solution to obtain all the benefits of D2D communications with the potential of reducing the communication latency, while also alleviating traffic on the already congested cellular bands~\cite{andreev2015understanding, liang2018cluster, bouzouita2019estimating}. 

While several commercial D2D communication radio access technologies (RATs), such as Bluetooth~\cite{bluetooth_spec}, ZigBee~\cite{zigbee_spec}, and Wi-Fi-Direct~\cite{wifidirect}, which operate on unlicensed frequency bands, already exist, and various other solutions to this end have been proposed in the past~\cite{dtvD2D, wifiD2D1, wifiD2D2, wifiD2D3, D2DU5G, mmwaveD2D2, D2Dwifi, D2Dcluster, mmwaveD2D1}, including the long term evolution (LTE) D2D by 3rd generation partnership project (3GPP)~\cite{lte_d2d_wp}, we identify the following major shortcomings in these prior-arts.

\textcolor{black}{First, many of the existing D2D solutions demand a dedicated radio in addition to the already existing cellular radio in the user equipment (UE). This results in increased hardware overhead and additional cost, which are undesirable, especially for low-cost and low-power MTC and IoT devices~\cite{liberg2018mtcbattery,BORKAR2020145}. The SL-U method proposed in~\cite{D2Dwifi} based on an optimal scheduling and resource allocation algorithm, the D2D clustering based technique in~\cite{D2Dcluster} that achieves SL communication using an SL device with the best channel interacting with the BS on behalf of every other user in its cluster, and the millimeter and microwave based D2D communication solution in~\cite{mmwaveD2D2}, all require Wi-Fi interfaces to enable SL operation.}

\textcolor{black}{Second, several of the prior arts require continued aid from the network BS for operation assistance and control, e.g., resource scheduling, which undermines the gains achieved by transitioning SL to unlicensed bands. The millimeter wave unlicensed band D2D communications solutions in~\cite{mmwaveD2D1, mmwaveD2D2}, the SL-U on television white space solution in~\cite{dtvD2D}, the D2D clustering based technique in~\cite{D2Dcluster}, and several other prior arts in~\cite{wifiD2D1, wifiD2D2, wifiD2D3, D2DU5G}, all rely on a central BS for scheduling, data collection and processing, and/or resource grant allocation.}


\textcolor{black}{Furthermore, without further enhancements, commercially available technologies such as Bluetooth and Wi-Fi-Direct require manual device pairing and repeated user interventions to re-establish connections when disruptions occur~\cite{bluetoothuser1,bluetoothuser2,bluetoothuser3}.}


In contrast to the prior-art solutions discussed above, we target a protocol which enables D2D communication on unlicensed bands in limited-capability MTC devices, such as those that use half-duplex and frequency division duplexing (HD-FDD) for cellular operation, which is a standardized design to reduce complexity in low-cost MTC~\cite{hoglund2018overview, BORKAR2020145, sequans_whitepaper,liberg2018mtcbattery, hdfdd_ref}. Additionally, our goal is also to develop a protocol that can operate with minimal guidance from a central BS and without any manual user intervention. To this end, we design a new protocol, called SCUBA, for Sidelink Communications on Unlicensed BAnds, which
\begin{itemize}
	\item is compatible with UEs that use HD-FDD operation,
	\item uses the existing single cellular radio to function in a time division multiplexed (TDM) manner as the secondary RAT on the device,
	\item meets typical SL latency and battery-life targets,
	\item conforms to regulatory requirements in various geographical regions, and
	\item provides flexibility for latency-power trade-off.
\end{itemize}
\textcolor{black}{While transitioning SL operation to the unlicensed bands and functioning with a single radio architecture solution enabled by the TDM operation reduce the monetary cost associated with SCUBA, our proposed protocol also presents a low opportunity cost by coexisting with the underlying primary RAT without requiring any modifications to the legacy cellular operation.}

In a precursor to this work~\cite{rajendran2020}, we presented a detailed feasibility analysis of an in-device TDM SL protocol for low-cost HD-FDD devices, where we identified unoccupied time-slots in cellular UEs to accommodate a secondary RAT for SL communications, and also specified requirements for the new SL-U protocol to meet regulatory requirements in the European and North American regions. Despite addressing these issues, we still face the following additional challenges for SL-U. Since we target deploying SL-U to operate on low-cost and low-complexity devices, such as MTC and narrowband IoT (NB-IoT) UEs, SL-U should ensure extended UE battery life while also providing reasonable control- and user-plane latency. Further, since we aim to operate independently without constant control and assistance from a central BS, SL-U must enable UEs to operate in a distributed manner, e.g., determining SL availability status of the destination UE. 

Considering the above design criteria and challenges, we develop SCUBA as an additional secondary RAT to cellular communication in a TDM fashion, such that it \textit{exploits the existing hardware} radio resource during idle times. 

\textcolor{black}{We summarize the contributions of this paper as follows:}
\begin{itemize}
    \item \textcolor{black}{We develop a novel medium access control (MAC) protocol and describe the key attributes and operation of SCUBA.}
	\item \textcolor{black}{Inspired by cellular discontinuous reception (DRX) operation, we introduce a customized SL-DRX mechanism for SCUBA to extend UE battery life while also providing a flexible tradeoff between latency and power consumption to serve diverse applications. We detail our proposed hybrid automatic repeat request procedure and the grant-based and non-grant based SCUBA operation in an SL connected mode.}
	\item \textcolor{black}{We design tailored adaptations for SCUBA to cater to special operating conditions, such as busy primary RAT traffic and latency critical SL applications.}
	\item \textcolor{black}{We provide comprehensive analyses and simulation results of performance evaluation of SCUBA including power consumption, network latency, and packet collisions, and compare the performance against existing commercial D2D solutions for a variety of MTC traffic conditions.}
\end{itemize}
\textit{Outline}: The rest of the paper is organized as follows. We present the preliminaries including the system and traffic models in Section~\ref{sec:system_model_prelims}, and the SCUBA design in Section~\ref{sec:d2dprotocol}. We provide power consumption and collision rates analyses in Section~\ref{section:analysis}, and their simulation results in Section~\ref{section:simulation}. In Section~\ref{section:discussion}, we discuss our protocol by highlighting its salient features and identifying potential future work. Finally, conclusions are drawn in Section~\ref{section:conclusion}. \textcolor{black}{We include the justification for the choice of specific simulation settings in the appendix.} Furthermore, for ease of reading, we have provided lists of acronyms and their definitions in Table~\ref{table:acronyms} and important notations and their meanings in Table~\ref{table:notations}.
\begin{table*}[t]
	\centering
	\caption{List of Acronyms} \label{table:acronyms}
	\begin{tabular}{ll|ll}
		\hline
		Acronyms  & Definitions                                 & Acronyms  & Definitions                                      \\ 
		\hline\hline
		3GPP      & 3rd Generation Partnership   Project        & NB-IoT    & Narrowband IoT                                    \\
		ACK       & Acknowledgment message                      & NR SL     & New Radio Sidelink                                \\
		API       & Application Program Interface               & PF        & Paging Frame                                      \\
		BLE       & Bluetooth Low Energy                        & PHY       & Physical Layer                                    \\
		BS        & Base Station                                & PO        & Paging Occasion                                   \\
		CCH       & Control CHannel                             & PRB       & Physical Resource Block                           \\
		CDRX      & Connected mode with discontinuous reception & RAI       & Release Assistance Indication                     \\
		ConA      & Connected Active mode                       & RAT       & Radio Access Technology                           \\
		CQI       & Channel Quality Indication                  & RLC       & Radio Link Control                                \\
		D2D       & Device-to-Device communication              & RRC       & Radio Resource Control                            \\
		Data-INAT & Data INActivity Timer                       & RX        & Receive                                           \\
		DL        & Downlink                                    & SAM       & Sidelink Availability Message                     \\
		DRX       & Discontinuous reception                     & SAM-D     & Sidelink Availability Message   for Dynamic SL-PO \\
		DRX-INAT  & DRX INActivity Timer                        & SAM-U     & Unavailability Sam                                \\
		DST       & Destination UE                              & SCH       & Shared CHannel                                    \\
		eDRX      & Extended DRX                                & SCUBA     & Sidelink Communications on   Unlicensed BAnds     \\
		ERP       & Effective Radiated Power                    & SCUBA-LLM & SCUBA Low Latency Mode                            \\
		FCC       & Federal Communications   Commission         & SF        & Subframe                                          \\
		GNSS      & Global Navigation Satellite   System        & SIB       & System Information Block                          \\
		HARQ      & Hybrid Automatic Repeat Request             & SL        & Sidelink                                          \\
		HD-FDD    & Half-Duplex and Frequency   Division Duplex & SL-DRX    & Sidelink-Discontinuous reception                  \\
		IAT       & Inter-Arrival Time                          & SL-INAT   & Sidelink inactivity timer                         \\
		IDRX      & Idle mode with discontinuous   reception    & SL-PO     & Sidelink paging occasion                          \\
		IoT       & Internet of Things                          & SL-U      & SL in Unlicensed frequency bands                  \\
		LTE       & Long Term Evolution                         & SRC       & Source UE                                         \\
		LTE-M     & LTE-MTC                                     & SW        & Switching SF                                      \\
		LUT       & Look-Up Table                                & TB        & Transport Block                                   \\
		MAC       & Medium Access Control                       & TBS       & Transport Block Size                              \\
		MCL       & Maximum Coupling Loss                       & TDM       & Time Division Multiplex                           \\
		MCS       & Modulation and Coding Scheme                & TX        & Transmit                                          \\
		MTC       & Machine Type Communication                  & UE        & User Equipment                                    \\
		NACK      & Negative acknowledgment message             & UL        & Uplink                                         \\
		\hline
	\end{tabular} 
\end{table*}
\begin{table*}[t]
	\centering
	\caption{List of important notations used in the paper}\label{table:notations}
	\begin{tabular}{ll|ll}
		\hline
		Notation              & Meaning                                       & Notation   & Meaning                                           \\ \hline \hline
		$\alpha_\ue$          & UE identity derived from IMSI                     & $N_\cdrx$   & Cellular CDRX cycle in units of SFs         \\ \hline
		$E_\nodata$           & Energy consumption during ﻿SCUBA idle state       & $N_\idrx$   & Cellular IDRX cycle in units of SFs         \\ \hline
		$E_\rxdata$           & Energy consumption during ﻿SCUBA reception        & $N_\sam$    & SAM period                                  \\ \hline
		$E_\sltx$             & Energy consumption during ﻿SCUBA transmission     & $N_\samu$   & SAM-U transmission interval in units of SFs \\ \hline
		$i_\slpo$             & SF index of the SL-PO                             & $N_\samd$   & SAM-D transmission interval in units of SFs \\ \hline
		$N_{\text{cluster}}$  & Number of clusters in SL-PO                       & $n_\samu$   & SAM-U duration in units of SFs              \\ \hline
		$n_{\text{dist}}$     & SF separation between clusters in SL-PO           & $n_\samd$   & SAM-D duration in units of SFs              \\ \hline
		$N_{\text{frame}}$    & SL-HARQ frame length in units of SFs              & $n_\slpo$   & SL-PO duration in units of SFs              \\ \hline
		$N_{\text{SL-grant}}$ & SF separation between SL data and SL grant        & $n_\slinat$ & SL inactivity time in units of SFs          \\ \hline
		$n_\sl$               & SL data duration in units of SFs                  & $N_\sldrx$  & SL-DRX cycle in units of SFs                \\ \hline
		$N_\harq$             & Number of ﻿parallel synchronous SL HARQ processes & $T_\sldrx$  & SL-DRX cycle in units of radio frames       \\ \hline
	\end{tabular}
\end{table*}
\section{Preliminaries}\label{sec:system_model_prelims}
\subsection{System Model}
\begin{figure}[t]
	\centering
	{\includegraphics[width=6cm]{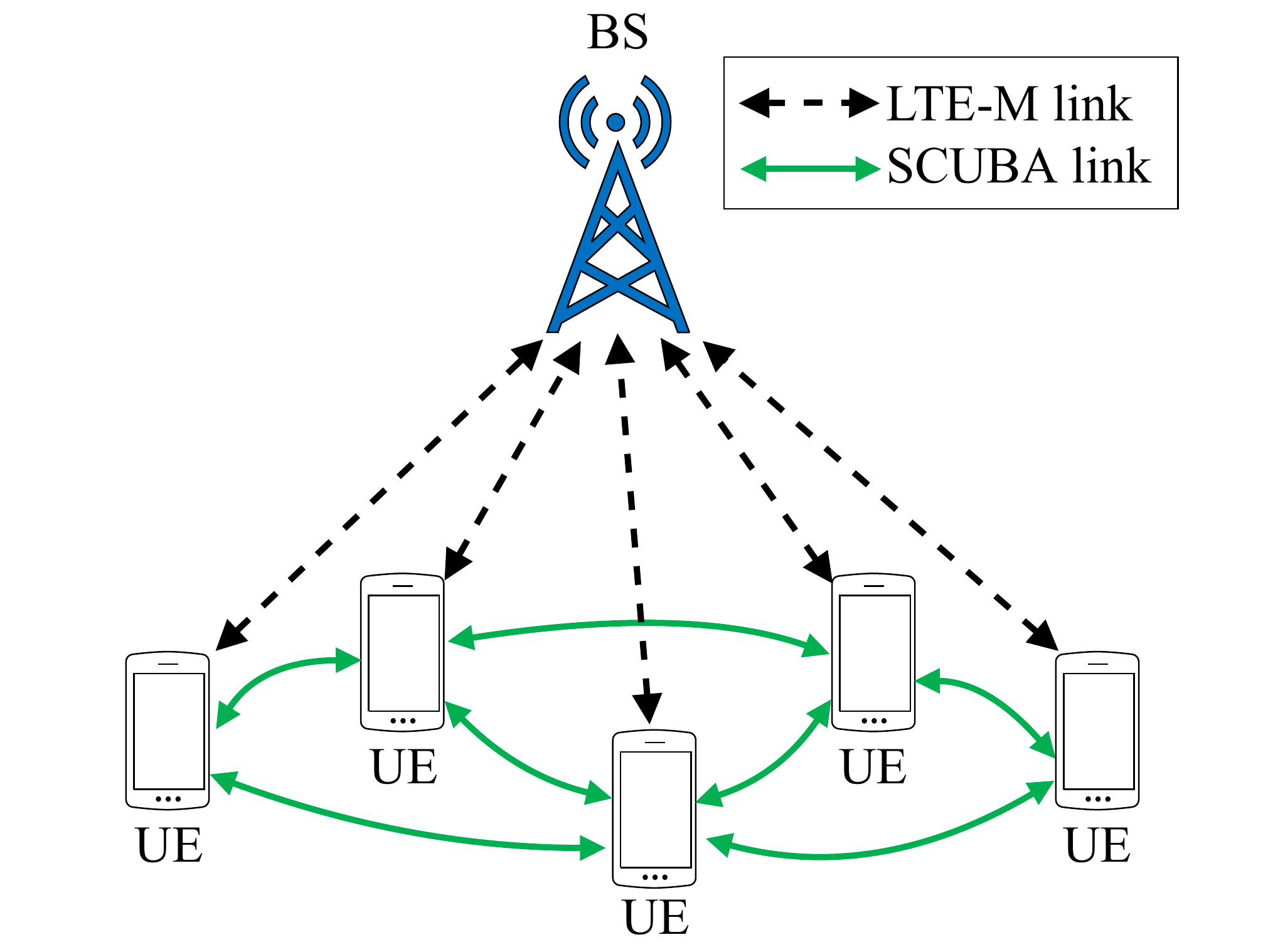}}
	\caption{A realization of SCUBA coexisting in an LTE-M network.}
	\label{fig:ueoperation} 
\end{figure}
We consider a hybrid network of UEs with SCUBA as the secondary RAT that is integrated with a primary \textit{LTE-like} RAT such as LTE, LTE-Advanced, LTE-MTC (LTE-M), or NB-IoT. In the following, we choose LTE-M operating on an elementary HD-FDD radio as the primary RAT for the design and analyses, such that constraints adopted in this study can be relaxed to extend the analysis for devices with greater capability, e.g., full-duplex FDD as in 5G new radio (NR) UEs. A conceptual representation of a SCUBA integrated network with LTE-M as the primary RAT and SCUBA enabling D2D on the unlicensed frequencies as a secondary RAT is shown in Fig.~\ref{fig:ueoperation}.
\subsubsection{UE operation in primary RAT}
We consider low-cost and low-complexity HD-FDD LTE category-M1 (Cat-M1) UEs~\cite{ts_36300}. Since we focus on the example of LTE-M being applied in the primary RAT, we follow the mode operations and frame structures accordingly. It can be seen that adopting the study to other primary RATs is straightforward. 

\begin{figure}[t]
	\centering
	{\includegraphics[width=7cm]{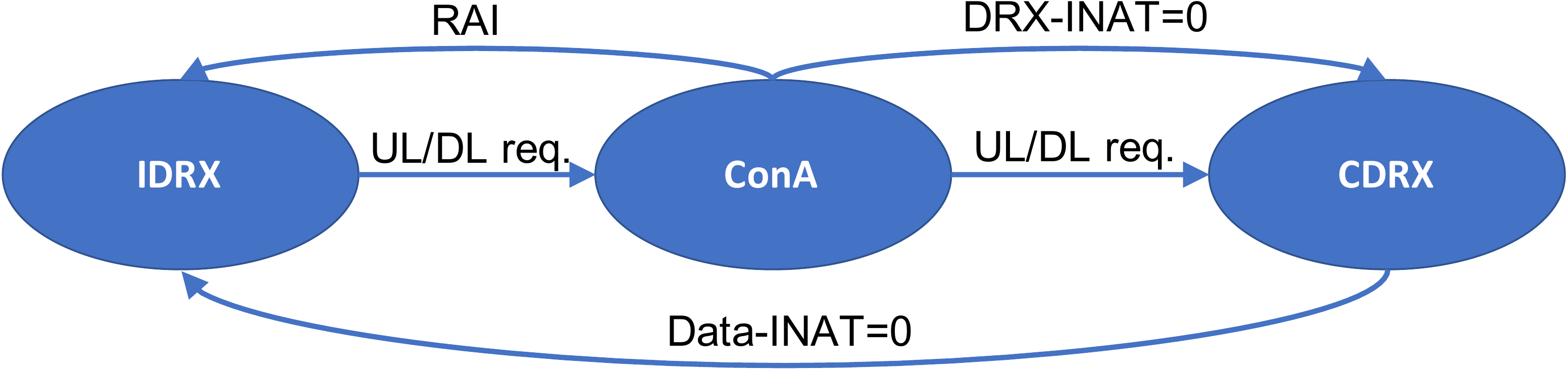}}
	\caption{State transition diagram for cellular operating modes.}
	\label{fig:uustates} 
\end{figure}

In time, the UL and DL messages are transmitted in $1$~ms long time units called subframes (SFs), with UL and DL SFs separated by guard periods for hardware switching between the receiver and transmitter chains. 
The UE operates in one of two states: radio resource control (RRC) connected or RRC idle. The UE enters the RRC connected state when it either receives data in DL or has data to transmit in UL. RRC connected state starts with a connected active mode (ConA), where the UE regularly communicates with the BS through UL and DL SFs with guard periods in between. Every data transmission triggers the restart of two timers namely, DRX inactivity timer (DRX-INAT)~\cite{koc2014device} and data inactivity timer (Data-INAT)~\cite{ts_36331}. After both the DL and UL data buffers are emptied, the UE enters a DRX inactivity period where it continuously monitors for DL messages, during which the DRX-INAT decrements. Upon the expiry of DRX-INAT, the UE enters the connected mode with discontinuous reception (CDRX), where it follows sleep cycles by waking up periodically to listen for DL only on CDRX-ON durations. If the UE receives DL data during the DRX inactivity period or CDRX-ON duration, it transitions back to the ConA mode. Else, the UE remains in the CDRX mode during which the Data-INAT decrements. Further upon the expiry of Data-INAT, the UE enters idle mode with discontinuous reception (IDRX) by releasing the RRC resources, where it undergoes longer sleep cycles by waking up periodically to page for DL only on IDRX paging occasions (POs). The UE goes back to the RRC connected state through an RRC connection setup procedure when it either receives valid paging in IDRX PO or has UL data to transmit. Additionally, when the UEs are provided with the release assistance indication (RAI) by the BS, they are allowed to skip the CDRX mode and transition directly from ConA to IDRX~\cite{ts_36321}. We summarize this operation in Fig.~\ref{fig:uustates}. 
\subsubsection{UE operation in secondary RAT}
\begin{figure}[t]
	\centering
	{\includegraphics[width=8cm]{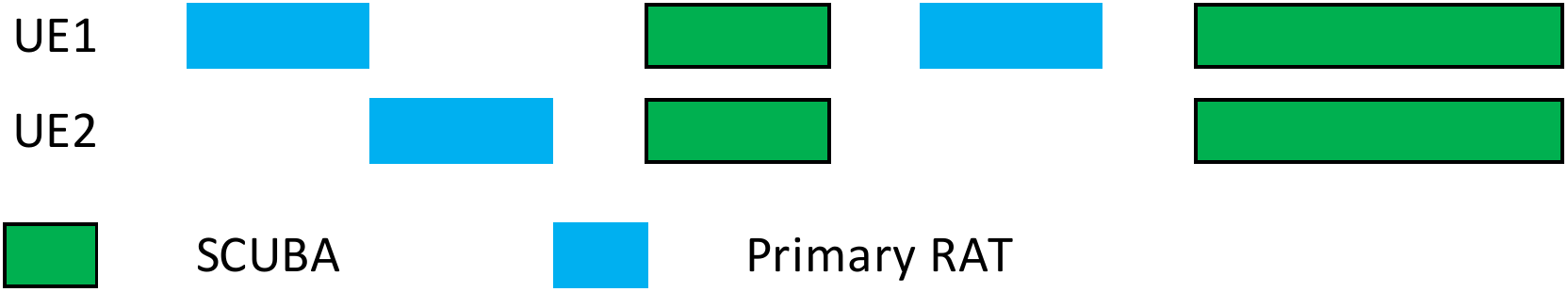}}
	\caption{An illustration of the TDM nature of SCUBA, which operates as a secondary RAT only when both devices are idle in their primary RATs.}
	\label{fig:tdmoperation} 
\end{figure}
SCUBA is a secondary RAT that allows UEs to communicate directly with each other in the SL. To this end, SCUBA ensures that it operates in a TDM manner with the primary RAT, such that the cellular communication, i.e., the primary RAT, operates unimpeded considering the HD-FDD constraint. An illustration of TDM operation of SCUBA along with primary RAT is shown in Fig.~\ref{fig:tdmoperation}, where the UEs communicate with each other via SCUBA only when both are free from their respective primary RAT. 

\begin{table}
\textcolor{black}{
	\caption{Regulatory requirements} \label{table:regulations}
	\begin{tabular}{c|c|c}
		\hline
		Parameters & Europe & US  \\ 
		&(865-868 MHz) & (902-928MHz) \\ \hline \hline
		Maximum ERP & 25 mW  & 1 W \\ \hline
		Maximum duty cycle & 1\%  & - \\ \hline
		Minimum bandwidth & - & 500 kHz\\ \hline
	\end{tabular}}
\end{table}

We design SCUBA to operate in the unlicensed bands of $865-868$~MHz in Europe and $902-928$~MHz in United States, as they are comparatively less occupied by other unlicensed RATs~\cite{rajendran2020}. We provide the summary of the relevant regulatory requirements, which govern the choice of effective radiated power (ERP), duty cycle, and bandwidth of SCUBA in Table~\ref{table:regulations}~\cite{rajendran2020}. Nevertheless, SCUBA is also applicable across other unlicensed spectra together with any applicable upper-layer traffic shaping that may be required to satisfy the duty cycle regulations associated with the used frequencies. These restrictions drive the physical layer design and the applicability of SCUBA across communication systems. For example, since the United States (US) Federal Communications Commission (FCC) regulates a minimum transmission bandwidth of $500$~kHz to be used in the $902-928$~MHz band, SCUBA may not be applicable in some devices which only support NB-IoT operating bandwidth of $180$~kHz.

\subsection{Traffic Model}
\begin{table}[!t]
	\centering
	\caption{Traffic Model~\cite[Annex. A]{tr_36888}} \label{table:trafficmodel}
	\begin{tabular}{c|c|c}
		\hline
		RAT                                                                                & Traffic Model & IAT (s)   \\ \hline \hline
		\multirow{2}{*}{\begin{tabular}[c]{@{}c@{}}Primary RAT\\ e.g., LTE-M\end{tabular}} & Periodic      & 300       \\ \cline{2-3} 
		& Poisson       & Mean = 30 \\ \hline
		SCUBA                                                                              & Poisson       & Mean = 30 \\ \hline
	\end{tabular}
\end{table}
We consider the MTC traffic models recommended by 3GPP~\cite[Annex. A]{tr_36888} for the design, analysis, and simulation of SCUBA by applying MTC traffic at both cellular link and SL. As shown in Table~\ref{table:trafficmodel}, we consider both types of traffic models with the respective inter-arrival times (IATs) as specified in~\cite[Annex. A]{tr_36888}, i.e., periodic and Poisson data arrivals for the cellular link. On the other hand, since the Poisson traffic model with a much smaller IAT simulates a busy traffic scenario, we apply this traffic type for SCUBA to evaluate its worst-case condition and guarantee performance targets.

\section{SCUBA}\label{sec:d2dprotocol}
\begin{figure}[t]
	\centering
	{\includegraphics[width=7cm]{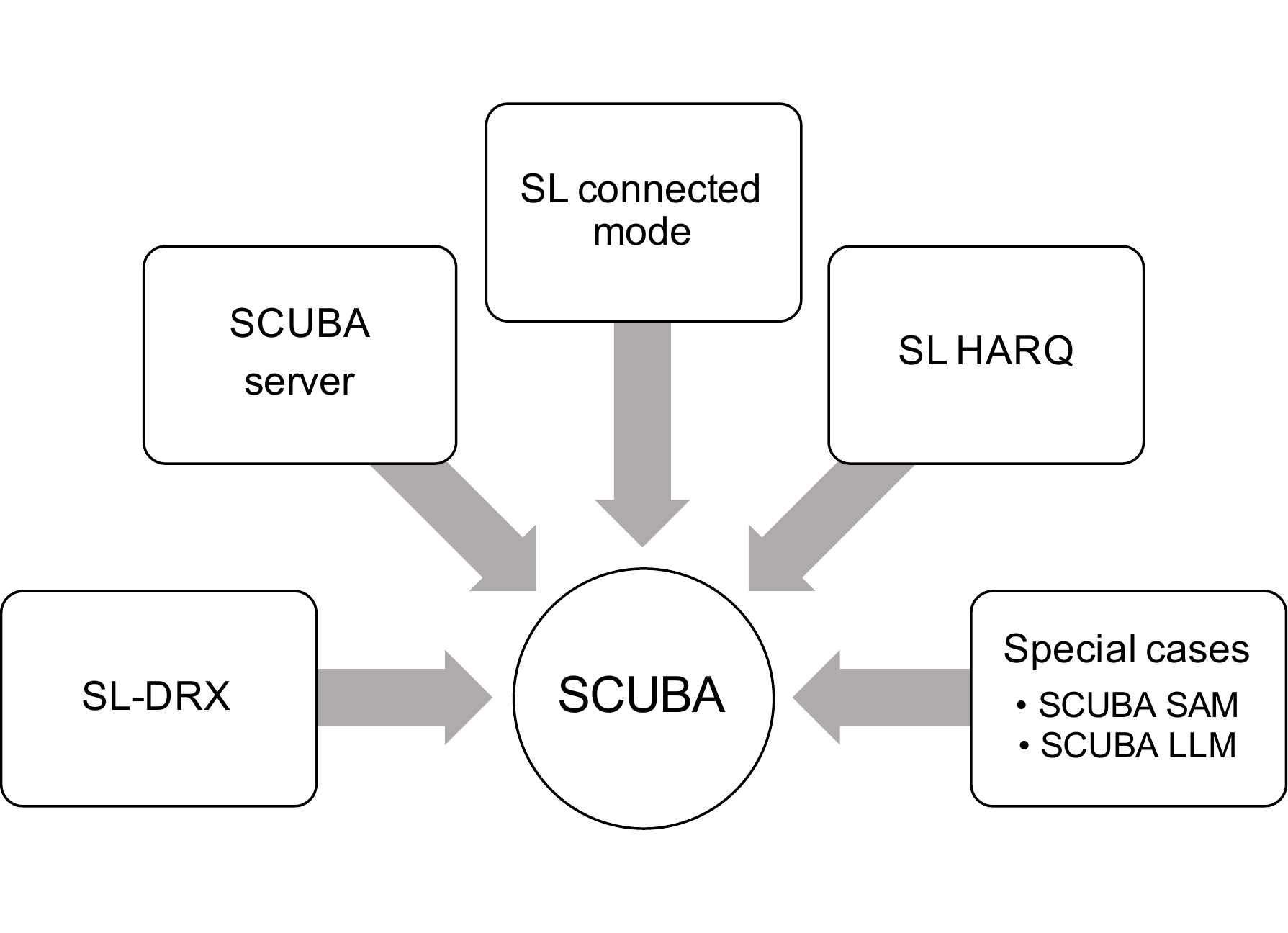}}
	\caption{Building blocks of SCUBA protocol.}
	\label{fig:SCUBAoverview} 
\end{figure}

\begin{table}[t]
	\centering
	\caption{Opportunities Available for SCUBA in Each Cellular Mode} \label{table:SL_opportunities}
	\begin{tabular}{c|c|c}
		\hline
		Cellular mode & Cellular Operation & Availability for SCUBA \tabularnewline
		\hline
		\hline
		ConA & DL/UL & No \tabularnewline
		& Switch SF & Yes \tabularnewline
		& DRX inactivity & No \tabularnewline
		& RRC connection setup & No \tabularnewline
		\hline
		CDRX & DRX-ON & No \tabularnewline
		& DRX-OFF & Yes \tabularnewline
		\hline
		IDRX & IDRX PO & No \tabularnewline
		& IDRX sleep & Yes \tabularnewline
		\hline
	\end{tabular} 
\end{table}
As SCUBA is intended to operate on a device that uses an \textit{LTE-like} protocol as the primary RAT, it borrows the physical layer (PHY) signal waveforms and also higher layer specifications, except the MAC layer, from the baseline protocol used within the UE. We thus focus on filling this gap by defining the MAC layer of SCUBA in this section. We define the MAC protocol by first introducing each of its fundamental structural components, as shown in  Fig.~\ref{fig:SCUBAoverview}, that finally integrate to provide the overall SCUBA MAC protocol. 

Prior to describing our protocol, we first highlight the challenges that we face and system requirements that need to be met by SCUBA. SCUBA is intended to provide an integrated single-radio SL-U solution for cellular UEs by incorporating a TDM SL as a secondary RAT that reuses the existing hardware. Therefore, the first challenge is to identify time occasions in the UE that are available for SCUBA transmission, i.e., the time instances where the UEs are free from cellular operations. For LTE-M as the primary RAT, a compiled list of the available opportunities in each of ConA, CDRX, and IDRX modes is shown in Table~\ref{table:SL_opportunities}. The UE is available for SCUBA transmission and reception in CDRX and IDRX when it is scheduled to \textit{sleep} for cellular communication. Therefore, the UE is free for SCUBA in all times except when it listens for a possible DL message during the CDRX-ON period or IDRX PO. Considering typical MTC traffic types~\cite{tr_36888}, these periods are sufficient for efficient SCUBA communications. Nevertheless, the switching SFs that are allocated to switch the UE hardware from the transmission to reception modes, several of which may be available when the UE is in the ConA mode, can also be used for short SCUBA message transmissions, since these durations are typically over allocated, i.e., most UEs can switch their hardware well within an entire SF~\cite{rajendran2020}. Once these available time slots are determined, a straightforward SL implementation is to let all network UEs listen for an SL message at all opportunities. Consider a portion of the network shown in Fig.~\ref{fig:ueoperation}, where a source UE (SRC) has SCUBA data to transmit to a destination UE (DST). A continuous listening technique is beneficial to achieve near-instant communication and therefore provides negligible control- and user-plane latency, since a DST is nearly guaranteed to be listening on a particular time-slot, given the sparse nature of MTC traffic in the cellular link~\cite{tr_36888}. However, listening for a SCUBA message on all available time slots significantly drains the UE battery. Therefore, to provide a flexible trade-off between the desired battery and latency, we borrow and extend the idea of DRX from cellular communications into SCUBA to introduce the concept of SL-DRX.

\subsection{SL-DRX}\label{section:sldrx}
Inspired by DRX modes present in conventional cellular communication~\cite{koc2014device, tseng2016delay}, we introduce SL-DRX cycles in SCUBA to provide a flexible application-controlled trade-off between network latency and UE battery life. As a solution, we specify SL paging occasions (SL-POs) during which the device wakes up to page for a potential SL message from an SRC. Every UE pages for a duration of $n_\slpo$ SFs once in every $N_\sldrx$ SFs, where $N_\sldrx$ is the SL-DRX cycle in units of subframes. We let the application program interface (API) choose $N_\sldrx$ according to the latency-power consumption trade-off it desires, including the option of setting $N_\sldrx=0$ for latency critical applications (see Section~\ref{section:llm}). 

As the number of UEs in the network grows, the probability of two or more UEs sharing the same SL-PO also increases, which results in an increased rate of packet collisions.
Recall that unlike other proposed SL-U solutions~\cite{dtvD2D, wifiD2D1, wifiD2D2, D2DU5G}, SCUBA does not rely on a central BS for resource scheduling or operation monitoring. Therefore, we introduce an inherent elementary collision control mechanism in choosing the SL-POs. The collision control method we propose reduces potential slot overlaps, i.e., overlap of SL-POs of different UEs and also an overlap of SL-POs with the IDRX PO of the UE primary RAT. 

We begin with a brief overview of selecting the IDRX-PO, which also guides us in designing the SL-PO slot allocation. The IDRX-PO location in conventional cellular operation is computed by first locating the paging frame (PF) as~\cite{ts_36304}
\begin{align}\nonumber
	&i_{\text{PF}} = i_{\text{SFN}}\in \mathcal{I}_{\text{SFN}}\\
	\text{s.t.}~~&i_{\text{SFN}}~\mathrm{mod}~T_\idrx=N_{\text{ID}},
\end{align}
where $\mathcal{I}_{\text{SFN}}$ is the set of system frame numbers (SFNs), whose range of values is $\mathcal{I}_{\text{SFN}}=\{0,1,..,1023\}$ for \textit{LTE-like} standards, $i_{\text{SFN}}$ is one SFN from the set, $i_{\text{PF}}$ are those SFNs which qualify to be the paging frames,
\begin{align}
N_{\text{ID}}&=\frac{T_\idrx}{N_\text{min}}(\alpha_\ue~\mathrm{mod}~N_\text{min}), \\
N_\text{min}&=\mathrm{min}(T_\idrx,~N_{\text{control}}), \\
\alpha_\ue&=\alpha_{\text{IMSI}}~\mathrm{mod}~\beta,
\end{align}
$T_\idrx$ is the IDRX paging cycle in radio frames, $N_{\text{control}}$ is a control parameter signaled in the system information block-2 (SIB2), $\alpha_{\text{IMSI}}$ is the international mobile subscriber identity (IMSI) of the UE, and the value of $\beta$ is $1024$~for LTE, $4096$~for NB-IoT, and $16384$~for NB-IoT on non-anchor carrier and LTE-M. Next, the pointing index $i_{\text{s}}$ that points to the exact location of the paging occasion inside an $i_{\text{PF}}$ is calculated as 
\begin{equation}\label{eq:is}
i_{\text{s}} = \Bigl\lfloor{\frac{\alpha_\ue}{N_\text{min}}}\Bigr\rfloor~\mathrm{mod}~\max \left(1,\frac{N_{\text{control}}}{T_\idrx}\right).
\end{equation}
Finally, the SF index of the PO, $i_{\text{PO}}$, inside the paging frame is obtained from a lookup table (LUT) based on the values of $i_{\text{s}}$ and $N_{\text{control}}$~\cite{ts_36304}.

\begin{figure}[t]	
	\begin{center}
		\subfloat[]{\includegraphics[clip,width=\columnwidth]{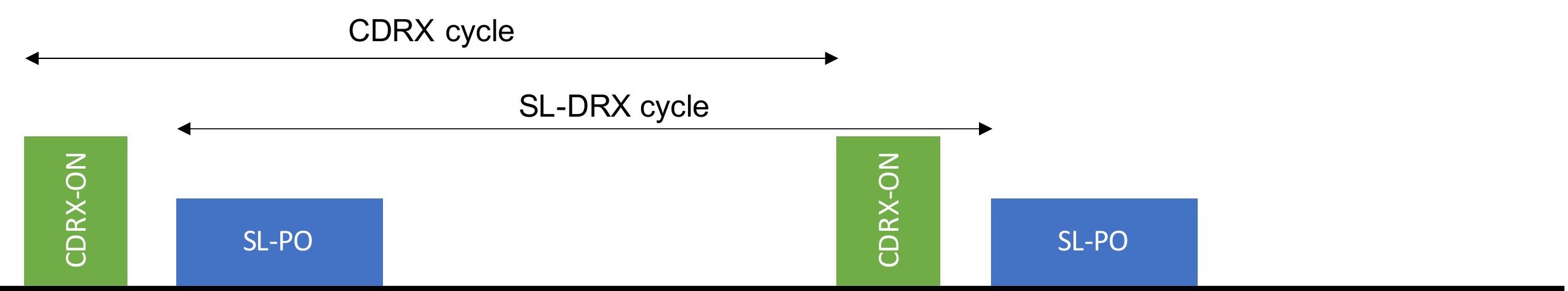}}	
		\\ \vspace{-.1in}
		\subfloat[]{%
			\includegraphics[clip,width=\columnwidth]{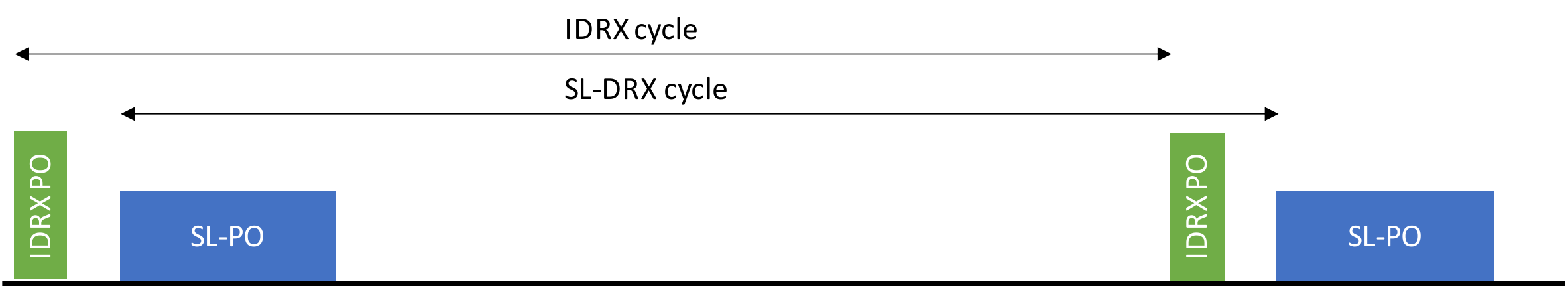}}
		\caption{SL-DRX cycles when the UE is in (a) CDRX mode (b) IDRX mode.}
		\label{fig:sldrx}
	\end{center}
\end{figure}

With this backdrop, we design the SL-PO for an SL paging cycle of $T_\sldrx$ radio frames, such that it does not overlap with $i_{\text{PO}}$. We compute $i_{\text{SL-PF}}$, the SL paging frame which accommodates the SL-PO, as
\begin{align}\nonumber
&i_{\text{SL-PF}} = i_{\text{SFN}}\in \mathcal{I}_{\text{SFN}}~
\text{s.t.}\\\label{eq:sl_po} 
i_{\text{SFN}}~\mathrm{mod}~T_\sldrx=&\begin{cases}
N_{\text{ID}}, \text{for}~T_\sldrx \ge T_\idrx\\
N_{\text{ID}}~\mathrm{mod}~T_\sldrx, \text{otherwise}.
\end{cases}
\end{align}
We then refer to the same LUT used for $i_{\text{PO}}$ to obtain the SF index of the SL-PO, $i_\slpo$, but offset it by $n_{\text{off}}$ SFs to assure no collision between IDRX PO and SL-PO. This UE IMSI-dependent SL-PO positioning randomizes the locations of the SL-PO and reduces the probability of inter-SL-PO overlaps. At the same time, an SRC can determine the SL-PO location of the DST using the a-priori DST IMSI information. We further propose a method to optionally modify the locations of SL-PO in Section~\ref{section:sam} to avoid the rare occurrence of an overlap of the SL-PO with ConA UL/DL communication and/or CDRX-ON duration.

For the paging cycles to be consistently periodic across different hyper-frames, which consists of $1024$ radio frames,~\eqref{eq:sl_po} requires $T_\sldrx$ to be a factor of the hyper-frame duration of 10.24 s, and thus SCUBA sets the allowed set of SL-DRX cycle values accordingly. Note that the same computation can be adapted when the primary RAT uses other forms of DRX, e.g., extended DRX (eDRX) as in LTE~\cite{ts_36304}, which can then also support SL-DRX cycles higher than $10.24$~s.


The SL-DRX cycles and SL-PO locations when the UE is in CDRX and IDRX are shown in Fig.~\ref{fig:sldrx}(a) and Fig.~\ref{fig:sldrx}(b), respectively. While IDRX PO is typically fixed to be a single SF, SCUBA provides a flexible SL-PO of $n_\slpo \geq 1$ to facilitate quick retransmissions. The multiple SFs in an SL-PO may be either continuous or interleaved to support efficient transmission of SL messages of different lengths. 

\begin{figure*}[t]	
	\begin{center}
		\subfloat[]{\includegraphics[clip,width=6.5in]{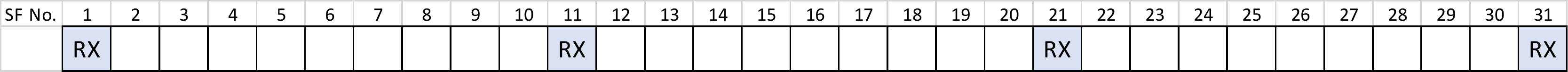}}	
		\\ 
		\subfloat[]{%
			\includegraphics[clip,width=6.5in]{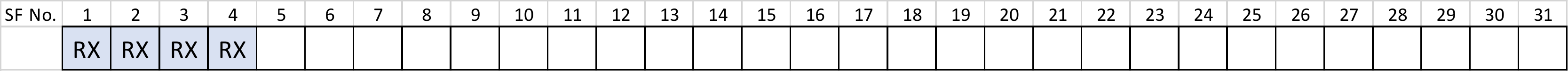}}
		\caption{Timing diagram for (a) Interleaved and (b) Consecutive SL-PO.}
		\label{fig:slpo}
	\end{center}
\end{figure*}

In general, we divide every SL-PO opportunity consisting of $n_\slpo$ SFs into $N_{\text{cluster}}$ separate clusters of SFs whose starting SFs are $n_{\text{dist}}$ SFs apart, with $n_{\text{dist}}N_{\text{cluster}} \geq {n_\slpo}$. Larger SL messages prefer higher values of $n_\slpo$ and lower $n_{\text{dist}}$ since multiple transport blocks can be transmitted with low latency, while smaller values of $N_{\text{cluster}}$ are desirable for short messages since the UE can go back to sleep without requiring to complete paging on all $n_\slpo$ SFs. At the same time, larger values of $n_{\text{dist}}$ also provide greater time diversity but with a possible increase in network latency. SCUBA provides UE APIs with the flexibility to choose these parameters based on the intended use-case. By carefully choosing desired values of parameters $\{n_\slpo, N_{\text{cluster}}, n_{\text{dist}}\}$, the resultant SL-PO may either be interleaved or consecutive. We show the examples of interleaved and consecutive SL-POs in Fig.~\ref{fig:slpo}(a) and (b) with $\{n_\slpo, N_{\text{cluster}}, n_{\text{dist}}\} = \{4, 4, 10\}, \{4, 1, \text{N/A}\}$, respectively, where the UE listens for paging on SFs marked as RX (receive).  


\subsection{SCUBA Server}\label{section:slserver}
As evident in Section~\ref{section:sldrx}, SCUBA requires an SRC to have {a-priori} knowledge of the $N_\sldrx$ and $\alpha_\ue$ of the DST to determine its $i_\slpo$ to transmit on. One solution to this end is to embed these parameters within every UE and update it as necessary when new UEs join the network. Such a method is suitable for a largely static network with few participating UEs. However, this technique introduces significant signaling overhead when the number of network nodes increases and/or UEs are mobile. Therefore, we propose the use of a central SCUBA server with the database of all UE and network parameters that can be accessed and updated using the conventional cellular link. A SCUBA server could be the network BS or a dedicated SCUBA infrastructure node. Since updating and accessing these parameters by UEs are typically infrequent ($\alpha_\ue$ is unique to every UE and hence is acquired only once per DST, and $N_\sldrx$ is fairly static), occasionally communicating with the SCUBA server introduces negligible signaling overhead.

\subsection{SL Connected Mode}\label{section:slconnectedmode}
\begin{figure}[t]
	\centering
	{\includegraphics[width=0.9\columnwidth]{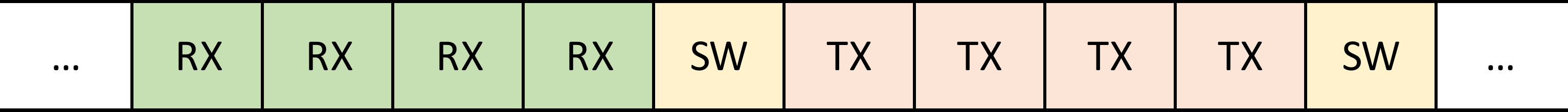}}
	\caption{An example of SL SF pattern for 4 parallel HARQ processes}
	\label{fig:slconnectedmode} 
\end{figure}

The latency associated with the SL communication is directly proportional to $N_\sldrx$ chosen by the DST. But after a SCUBA message exchange is initiated, waiting for the next SL-PO to continue transmission introduces unnecessary delays. Therefore, SCUBA includes an SL connected mode, where the SRC and DST UEs engage in interactive communication until both their associated SL buffers are emptied. Note that this operation continues to be contingent on the UE being free from its primary RAT, as is the case with all SCUBA operations. SL connected mode includes $N_\harq$ continuous transmit (TX) SFs followed by a switching (SW) SF and additional $N_\harq$ RX SFs and so on, with as many cycles as required for data transfer to complete. An example of the RX-TX SF pattern with $N_\harq=4$  is shown in Fig.~\ref{fig:slconnectedmode}. Following completion, we allot an SL inactivity time, controlled by the SL inactivity timer (SL-INAT), during which both UEs monitor for a potential SL message. This period potentially provides further reduction in network latency, similar to the DRX inactivity time seen in LTE cellular link.  

\subsection{SL HARQ}\label{section:harq}
To counter varying channel and noise conditions, we incorporate adaptive modulation and coding scheme (MCS) and backward error correction in the form of hybrid automatic repeat request (HARQ) in SCUBA. We support $N_\harq$ parallel synchronous HARQ processes in the SL connected mode to efficiently utilize the processing time of a transport block (TB) for reception of further TBs. Additionally, we also propose the use of the following two HARQ schemes in SCUBA which the UE API can choose from, along with a suitable value for $N_\harq$.

\subsubsection{Fixed MCS/TBS} 
In this method, the SRC transmits SL data and control channels using a fixed MCS and TB size (TBS) chosen from a small set of pre-defined values. The grant-free nature of this scheme results in blind decoding at the DST, which adds to the processing complexity and time, but improves latency as signaling overhead introduced by grants are avoided. The pre-defined set of MCS and TBS values can either be pre-programmed on all UEs, or can be obtained from the SCUBA server. After successfully decoding every SL TB, the DST responds to the SRC individually with an acknowledgment message (ACK) $N_\harq$ SFs after receiving the message. Therefore, the value of $N_\harq$ must also be chosen based on the computational ability of the UEs. The SRC retransmits any unacknowledged TB using the same pre-defined MCS and TBS, with the associated sequence number in the radio link control (RLC) header distinguishing a retransmission from a fresh transmission, similar to the use of RLC sequence number in LTE~\cite{ts_36322}. 


\subsubsection{Grant-based HARQ}
\begin{figure*}[t]	
	\begin{center}
		\subfloat[]{\includegraphics[clip,width=6.5in]{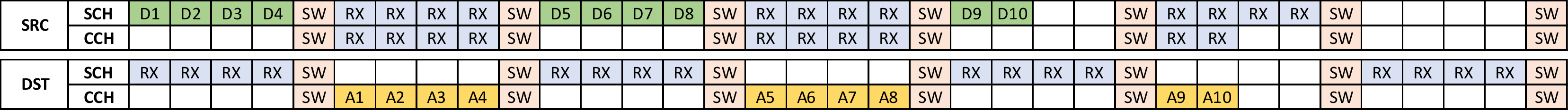}}	
		\\ 
		\subfloat[]{%
			\includegraphics[clip,width=6.5in]{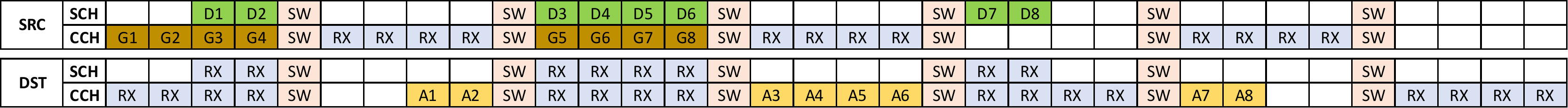}}
		\caption{Timing diagram for (a) Fixed MCS and (b) Grant-based HARQ schemes.}
		\label{fig:harqtimingdiagram}
	\end{center}
\end{figure*}
In this scheme, SRC begins transmission by sending a grant in the control channel for the upcoming data transmission, indicating its MCS and retransmission flag. The MCS is chosen adaptively using the channel quality indication (CQI) feedback provided by the DST, or with a conservative MCS for a cold transmission. Although this scheme may increase network latency due to the signaling overhead introduced by transmitting a grant preceding every data SF, it provides the ability to adapt transmission to the varying channel and noise conditions to maximize the overall throughput. To ease grant decoding complexity at the DST, we impose the SL grant to be transmitted at least $N_{\text{SL-grant}}$ SFs before the SL data is transmitted.



For defining grant and data locations, we define an SL-HARQ frame length as, $N_{\text{frame}} = 2(N_\harq + 1)$, which is the time taken to complete one cycle of TX-RX together with a SF for switching between TX and RX. Fig.~\ref{fig:harqtimingdiagram} shows the timing diagrams for the fixed MCS scheme (Fig.~\ref{fig:harqtimingdiagram}(a)) with $N_\harq = 4$, and the grant-based HARQ (Fig.~\ref{fig:harqtimingdiagram}(b)) with $N_{\text{SL-grant}}=2$ and $N_\harq = 4$. Similar to LTE, we use shared channel (SCH) to transmit the SL data, and control channel (CCH) to send grant and ACK. The notations G$i$, D$i$, and A$i$ in the figures represent the $i$th grant, data, and ACK, respectively. For a grant transmitted on the $i$th SF in an SL-HARQ frame, the corresponding SL data is transmitted on the $k$th SF, where 
\begin{align}\label{eq:tau_limits}
k =
\begin{cases}
i+N_{\text{SL-grant}},&\text{for}~i< N_\harq-1\\
i+2N_\harq,&\text{otherwise}.
\end{cases}
\end{align}

\subsection{Special Cases}\label{section:specialcases}
In the following, we describe two modifications to native SCUBA to accommodate special needs of dedicated use-cases. 
\subsubsection{SCUBA-SAM}\label{section:sam}
\begin{figure}[t]	
	\begin{center}
		\subfloat[]{\includegraphics[clip,width=\columnwidth]{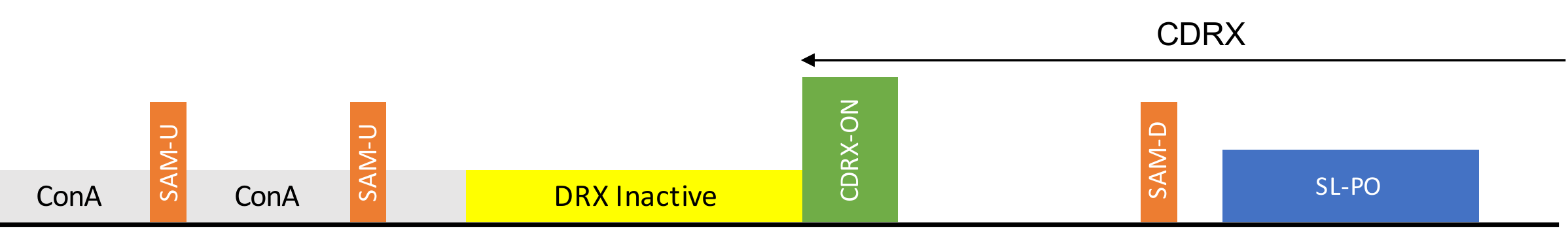}}	
		\\ 
		\subfloat[]{%
			\includegraphics[clip,width=\columnwidth]{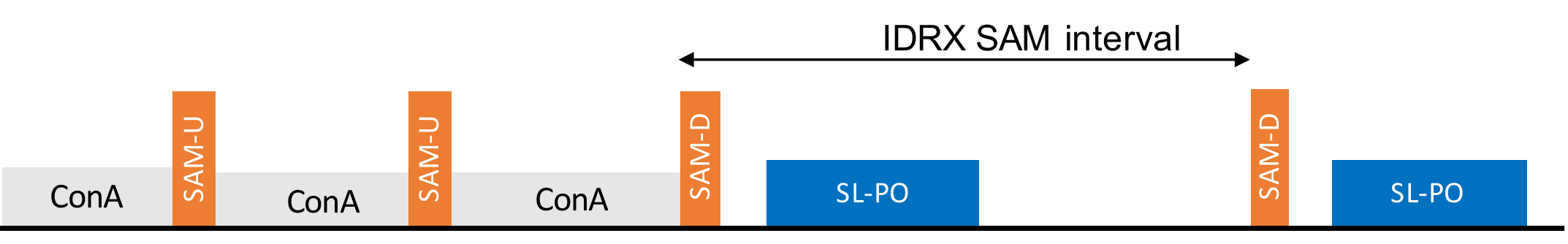}}
		\caption{SAM-D transmissions and the associated dynamic SL-POs when the UE enters (a) CDRX mode, and (b) IDRX mode directly, when RAI is activated.}
		\label{fig:samd}
	\end{center}
\end{figure}

\begin{figure}[t]
	\centering
	{\includegraphics[width=\columnwidth]{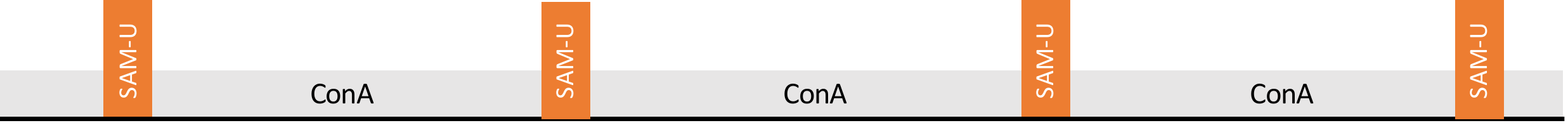}}
	\caption{SAM-U transmissions when the UE is in ConA.}
	\label{fig:samu} \vspace{-0.1in}
\end{figure}

When a UE is relatively busy with the primary RAT characterized by long durations of ConA, several of its SL-PO opportunities are overlapped by the cellular link, which in turn may lead to a massive increase in the control plane latency for SCUBA. This situation can be exacerbated when the DST uses a large SL-DRX cycle. To counter this problem, in addition to the existing SL-POs, we introduce dynamic SL-POs, which are positioned at the earliest available time opportunity when the UE is free from the cellular link. Such opportunities could either be the CDRX-OFF states, or the IDRX sleep duration when RAI is enabled. Since the locations of these dynamic POs are stochastic in nature and are unknown to other UEs, they need to be specifically advertised. We thus propose a short broadcast message called SL Availability Message for Dynamic SL-PO (SAM-D), as shown in Fig.~\ref{fig:samd}, which notifies UEs of an upcoming dynamic SL-PO. Any SRC can then decode a SAM-D to identify the SL availability of a DST. For a rational SAM scheme that does not introduce extended SAM-listening periods, both the availability and unavailability statuses of a UE needs to be regularly communicated. Accordingly, we also use SAM to indicate the unavailability of a DST UE, e.g., during the ConA mode. To this end, we utilize the short time opportunities available during the ConA, e.g., radio switching SFs and/or other idle SFs~\cite{rajendran2020}, to transmit an unavailability SAM (SAM-U), as shown in Fig.~\ref{fig:samu}. We set the length of SAM to be $n_\samd = n_\samu \leq 0.5$~SF to ensure that it fits within the shortest available time-frame in the ConA mode ($\approx 0.6$~ms)~\cite{rajendran2020}. Since continuous listening for a SAM introduces unreasonable power consumption, we define a SAM period of $N_\sam$ SFs within which every UE must transmit a SAM, and whose value can be obtained from the SCUBA server. We choose $N_\sam >$ DRX-INAT to guarantee that at least one SAM can be sent by every UE within $N_\sam$. Note that since SAM transmission and listening results in increased power consumption, we recommend SCUBA-SAM to be used only by UEs with busy cellular traffic, where ConA operation often blocks the SL-POs.



\subsubsection{SCUBA-LLM}\label{section:llm}
UEs that are not limited by battery life, e.g., alternating current (AC) powered devices, and are used for latency critical applications can use a special option of our protocol, called SCUBA low latency mode (SCUBA-LLM). This setting assigns $N_\sldrx=0$, and thus the UE listens for a potential SL message on all available time opportunities (see Table~\ref{table:SL_opportunities}) without entering the sleep state.  SCUBA-LLM subsumes SCUBA-SAM to enable SAM-U and SAM-D advertisements for control plane latency enhancement in busy devices, since the additional power required to transmit and receive SAMs is not a factor of consideration to these UEs that are not constrained by power consumption.

\subsection{Data Transfer}\label{section:sltx}
We now specify the rules for SCUBA transmission considering the features introduced in Sections \ref{section:sldrx} to \ref{section:specialcases}. Since SCUBA is a secondary RAT, its data transmission is always contingent on both the UEs, SRC and DST, being free from cellular operation. However, SCUBA does not impose any further traffic shaping measures as we already showed previously that, with the right choice of TBS, all types of MTC traffic specified by 3GPP~\cite{tr_36888} conforms well within the duty cycle limits introduced by regulatory authorities~\cite{rajendran2020}. 

\subsubsection{SCUBA Transmission} 
An SRC that receives transmission request from the application layer uses the values of $N_\sldrx$ and $\alpha_\ue$ (either obtained newly from the SCUBA server or that it knows a-priori) to compute $i_\slpo$ of the DST. Based on the HARQ method used and the SL-PO type applied at the DST, the SRC initiates transmission on the DST SL-PO. Upon positive reaffirmation of a successful transmission by way of ACK reception, the SRC enters the SL connected mode with the DST, and exits after SL data transfer completion. If no ACK is received, the SRC retries at the next available DST SL-PO. This operation is summarized in Algorithm~\ref{alg:only_alg}.

\begin{algorithm}[t]
	\centering
	\caption{SCUBA transmission procedure at the SRC.} \label{alg:only_alg}
	\begin{algorithmic}[1]
		\State \textbf{Start}: SL transmission request from the application layer
		\State Compute $i_\slpo$ of DST using~\eqref{eq:sl_po} and the LUT from~\cite{ts_36304}
		\State \textbf{while} SRC free from cellular link \textbf{do}
		\State Transmit grant/data on the next $i_\slpo$ of DST
		\If {ACK received}
			\State Enter SL connected mode
			\If {SL-INAT expires}
				\State \textbf{break}
			\EndIf
		\Else
			\State \textbf{go to}: Step 3
		\EndIf	
	\end{algorithmic} 
\end{algorithm}

\subsubsection{SCUBA-SAM Transmission}\label{sec:scuba_sam_tx}
\begin{figure}[t]
	\centering
	{\includegraphics[width=\columnwidth]{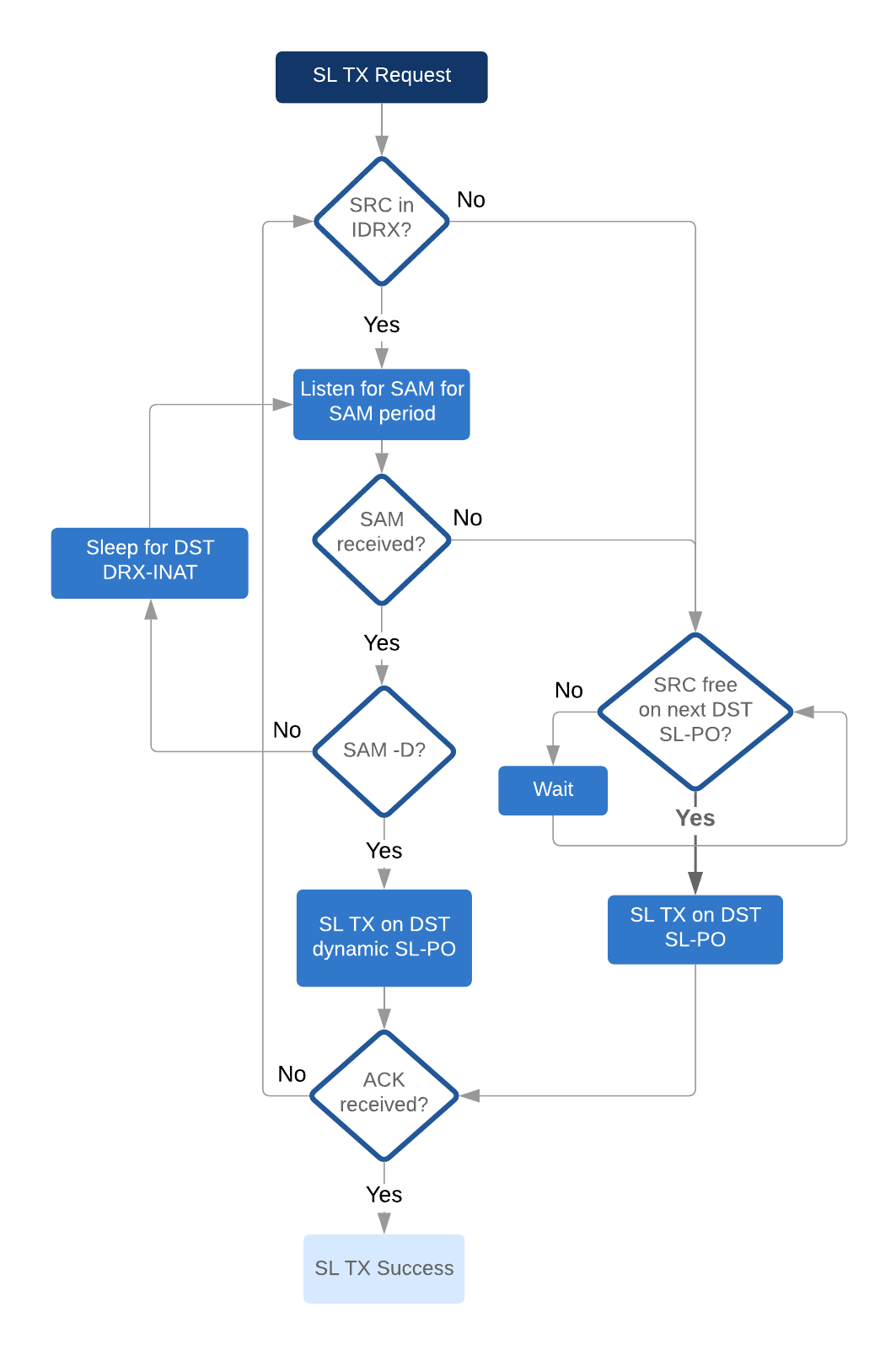}}
	\caption{Flowchart for SCUBA-SAM transmission.}
	\label{fig:flowchartsam} 
\end{figure}

Upon receiving an SL transmission request from the application layer, an SRC UE begins listening for a SAM for $N_\sam$ SFs as long as the SRC is not in the connected mode. If it receives a SAM-U from the DST indicating that the DST is in ConA mode, the SRC goes to sleep for a period of DRX-INAT, since that is the minimum period that the DST would remain in ConA. If the SRC receives a SAM-D instead, it initiates SL communication with the DST on the dynamic SL-PO indicated in the SAM-D. On the other hand, if the SRC does not receive any SAM, it concludes that the DST is in the IDRX state. It then computes the SL-PO of the DST and attempts transmission on $i_\slpo$. A successful SL initiation is indicated by an ACK transmitted by the DST. If the SRC does not receive an ACK, it retries transmission in the next $i_\slpo$. A summary of the SCUBA-SAM transmission is also provided in Fig.~\ref{fig:flowchartsam}.

\subsubsection{SCUBA-LLM Transmission}
The transmission procedure in LLM follows the same rules as in Section~\ref{sec:scuba_sam_tx} and Fig.~\ref{fig:flowchartsam}. When the DST also applies SCUBA-LLM, the DST SL-PO spans a large duration, which provides near-instant transmission opportunity to the SRC.

\section{Performance Analysis}\label{section:analysis}
In this section, we analyse the performance of our proposed protocol in terms of the power consumed in each UE and the packet collision rate in a SCUBA network for varying traffic conditions. Throughout the analysis, we consider fixed MCS/TBS scheme as defined in Section~\ref{section:harq} since conventional SL communication protocols like LTE-D2D are typically grant-free in nature. The analysis can be extended in a similar way for the grant-based HARQ scheme by making appropriate changes to include SL grant SFs.
\subsection{Power Consumption}
We analyze the power consumption for all three modes of operation. In general, the average overall power consumption can be computed as 
\begin{equation}
	P = \lambda_\txdata E_\txdata + \lambda_\rxdata E_\rxdata + \lambda_\nodata E_\nodata,
\end{equation}
where $\lambda_\txdata$, $\lambda_\rxdata$, and $\lambda_\nodata$ are the average event rates for UE having SL data to transmit, to receive, or neither, respectively, and $\lambda_\txdata + \lambda_\rxdata + \lambda_\nodata = 1$. $E_\txdata$, $E_\rxdata$, and $E_\nodata$ are the total energy consumed by the UE during SCUBA transmission, reception, and idle states, respectively. We begin with computing the energy consumption for completing the SCUBA transmission and reception, $E_\sltx$ and $E_\slrx$, respectively, for variable packet sizes as
\begin{align}\nonumber
	E_\sltx &= t_{\text{SF}} \Bigg( \underbrace{P_\tx n_\sl}_{\text{TX SL-U data}} + \underbrace{P_\rx n_\sl}_{\text{RX ACK}} 
		+ \underbrace{P_\switch \left\lceil \frac{n_\sl}{N_\harq}+1 \right\rceil }_{\text{Switch b/w TX and RX}}  \\ 
		&+ \underbrace{P_\rx n_\slinat}_{\text{SL inactivity}} \Bigg), \\\nonumber
	E_\slrx &= t_{\text{SF}} \Bigg(  \underbrace{P_\rx(N_\harq+\max({0, n_\sl-N_\harq}))}_{\text{RX SL data}} + \underbrace{P_\tx n_\sl}_{\text{TX ACK}} \\&+ \underbrace{P_\switch \left\lceil \frac{n_\sl}{N_\harq}+1 \right\rceil }_{\text{Switch b/w TX and RX}}  
	+ \underbrace{P_\rx n_\slinat}_{\text{SL inactivity}} \Bigg),	
\end{align}
where $P_\tx$, $P_\rx$, and $P_\switch$ represent the UE power consumption for any data TX, RX, and switching between TX and RX operations, respectively,  $n_\sl$ and $n_\slinat$ are the number of SFs consumed by SL data and SL inactivity time, and $t_{\text{SF}}$ is the time duration of a SCUBA SF.
Next, we compute $E_\txdata$, $E_\rxdata$, and $E_\nodata$  for each of three modes of operation. 
\subsubsection{Native SCUBA}
Under the native SCUBA operation, UEs only TX, RX, and listen for SL-POs during primary RAT inactivity. Therefore,
\begin{align}
	E_\txdata &= (1-p(\cona))E_\sltx, \\
	E_\rxdata &= (1-p(\cona))E_\slrx, \\
	E_\nodata &= (1-p(\cona))\frac{P_\rx n_\slpo t_{\text{SF}}}{N_\sldrx},
\end{align}
where $p(\cona)$, $p(\cdrx)$, and $p(\idrx)$ represent the probabilities of the UE to be in ConA, CDRX, and IDRX modes, respectively, in the primary RAT.
\subsubsection{SCUBA-SAM}
When UEs use the SAM mode, they additionally also transmit SAM in the connected modes to indicate their availability status. Since the use of SAM-D is limited in IDRX mode, its probability is negligible and hence not considered in the analysis. Thus,
\begin{align} \nonumber
&E_\txdata  =  p(\cona) \frac{P_\tx n_{\samu} t_\text{SF}}{N_{\samu}}  +  (1-p(\cona))\bigg(\!E_\sltx\bigg) \\ \nonumber
& + p(\idrx) \bigg( P_\rx t_{\text{SF}}\Big( p(\cona)\frac{k_{\samu}N_\samu + N_\samd}{2} \\ \label{eq:sltx_sam}
& + p(\cdrx) \frac{N_\samd}{2} + p(\idrx)N_\sam \Big) \bigg), \\ \nonumber
&E_\rxdata = p(\cona) \frac{P_\tx n_{\samu} t_\text{SF}}{N_{\samu}} \\ \label{eq:slrx_sam}
&+ (1-p(\cona))E_\slrx + p(\cdrx)\frac{P_\tx n_{\samd} t_\text{SF}}{N_\samd}, \\ \nonumber
&E_\nodata = p(\cona) \frac{P_\tx n_{\samu} t_\text{SF}}{N_{\samu}} \\ \nonumber 
&+ p(\cdrx) \left( \frac{P_\tx n_{\samd} t_\text{SF}}{N_{\samd}} + \frac{P_\rx n_\slpo t_{\text{SF}}}{N_\sldrx} \right) \\  
&+ p(\idrx) \frac{P_\rx n_\slpo t_{\text{SF}}}{N_\sldrx},
\end{align}
where $k_{\samu}$ is the average number of SAM-Us received by the SRC UE before receiving a SAM-D. $N_\samu$ and $N_\samd$ are the transmission intervals of SAM-U and SAM-D in SFs units, respectively, with preferably $N_\samd, N_\samu \leq \frac{N_\sam}{2}$ which guarantees SAM-D and SAM-U to be received by the DST listening for $N_\sam$ SFs.
\subsubsection{SCUBA-LLM}
The energy consumption in LLM is similar to that of SCUBA-SAM with the exception that the SL-POs extend throughout all available times in CDRX and IDRX modes, which impacts the power consumption when the UE is idle. Therefore, the transmission and reception energies are identical to~\eqref{eq:sltx_sam} and~\eqref{eq:slrx_sam}, respectively, and
\begin{align}\nonumber 
	&E_\nodata = p(\cona) \frac{P_\tx n_{\samu} t_\text{SF}}{N_{\samu}} \\ \nonumber
			  &+ {p(\cdrx)} \left(\frac{P_\tx n_{\samd} t_\text{SF}}{N_{\samd}} + \frac{P_\rx n_{\cdrx, \slpo, \text{LLM}}t_\text{SF}}{N_\cdrx} \right) 	 \\
			  &+ p(\idrx) \frac{P_\rx n_{\idrx, \slpo, \text{LLM}}t_\text{SF}}{N_\idrx},
\end{align}
where $N_\cdrx$ and $N_\idrx$ are the CDRX and IDRX cycle values in SFs of the primary RAT in the UE, and $n_{\cdrx,\slpo, \text{LLM}}$ and $n_{\idrx, \slpo, \text{LLM}}$ are the free time duration in SFs in CDRX and IDRX modes, respectively, which are all used for paging SL messages. 

\subsection{Packet Collisions}\label{section:packetcollisions}
SCUBA inherently includes an elementary collision control strategy by incorporating the SL-POs to be dependent on $\alpha_\ue$. Nevertheless, packets can still collide when two or more SRC UEs transmit data simultaneously to the same UE or to different UEs sharing overlapping SL-POs and SL bands. To compute the probability of SL data collision, we define the following events:
\begin{itemize}
	\item A: two or more SRC UEs have data to transmit at an SL-PO
	\item B: two or more SRC UEs transmit at the same SL-PO
	\item C: SRC and DST UEs are not in ConA.
\end{itemize}
The probability of collision can then be expressed as 
\begin{equation}
	p_\text{c} = \frac{1}{N_{\text{B}}} p(A \cap B \cap C),
\end{equation}
where ${N_{\text{B}}}$ is the number of orthogonal unlicensed frequency bands that can be used, depending on the total available bandwidth and number of physical resource blocks (PRBs) applied for SCUBA transmission. 
To evaluate the worst-case collision scenario, we set $p(C) = 1$ to emulate conditions where UEs are mostly free from cellular communication and available for SL-U\footnote{We have shown previously that Event C is overwhelmingly the most probable scenario encountered with typical MTC traffic~\cite{rajendran2020}.}. Therefore, we can express
\begin{equation}
	p_\text{c} = \frac{1}{N_{\text{B}}} p(A)p(B|A),
\end{equation}
with
\begin{equation}\label{eq:collision_pa}
	p(A) = \sum \limits_{k=2}^{N_\ue} \binom{N_\ue}{k}p_\sltx^k(1-p_\sltx)^{N_\ue - k},
\end{equation}
where $N_\ue$ is the total number of UEs in the SCUBA network and $p_\sltx$ is the probability of a UE having a non-empty SL buffer. The latter can be computed as
\begin{equation}
	p_\sltx =  \sum \limits_{n=1}^{\infty}\frac{\e^{-\gamma N_\sldrx t_\text{SF}}(\gamma N_\sldrx t_\text{SF})^n }{n!},
\end{equation}
where $\frac{1}{\gamma}$ is the mean inter-arrival time of the Poisson packet arrivals in SCUBA.
Furthermore,
\begin{equation}
	p(B|A) = \sum \limits_{k=2}^{N_\ue} \left(\frac{N_\slpo}{N_\sldrx}\right)^k
\end{equation}
under the assumption that an SRC UE transmits on an SL-PO with uniform random probability and that SL-POs are allotted in slots with the same SL-DRX cycles used by all UEs. The above analysis considers the case of consecutive SF SL-POs as shown in Fig.~\ref{fig:slpo}(b), but similar analysis also follows for the case where SL-POs are interleaved. Note that for applications where UEs frequently communicate with a central receiver, e.g., a data aggregator, we have $p(B|A)=1$ and therefore, $p_\text{c}=\frac{1}{N_{\text{B}}}p(A)$. We show results for $p_\text{c}$ in Section~\ref{section:simulation} across varying number of network participants that will guide the choice of SL-DRX cycle based on the SL use-case and the traffic being served. 

Similar to SCUBA data collisions, the SAMs transmitted as part of the UE state discovery procedure may also collide leading to a false positive for UE availability. Determining the SAM collision probability across various network conditions contributes to the choice of SAM transmission intervals along with the acceptable latency and power consumption trade-off. The SAM collision probability can be expressed similar to~\eqref{eq:collision_pa} as
\begin{equation}
	p_{\text{c,}\sam} = \frac{1}{N_{\text{B}}} \sum \limits_{k=2}^{N_\ue} \binom{N_\ue}{k}p_\sam^k(1-p_\sam)^{N_\ue - k},
\end{equation}
where $p_\sam$ is the SAM transmission probability computed as 
\begin{equation}\label{eq:sam_prob}
	p_\sam = p_\cona \frac{n_\samu}{N_\samu} + p_\cdrx \frac{n_\samd}{N_\samd}.
\end{equation}
The SAM collision rates along with the other performance indicators of SCUBA are presented in the following section.

\section{Evaluation}\label{section:simulation}
We evaluate the performance of SCUBA by simulating its MAC layer timing behavior in MATLAB. Along with power consumption and collision rates, we also use network latency as a performance indicator for evaluation. We define network latency as the time taken to complete an acknowledged transmission of $100$~bytes of SL data, which is the typical data size considered for triggered as well as periodic reports in MTC smart metering applications~\cite{tr_36888, smartmeter}. \textcolor{black}{We also include a comparison of the power consumption of native SCUBA against that of other D2D solutions for different sleep cycles.}
\subsection{Simulation Settings}
For the underlying primary RAT, we simulate both Poisson and periodic cellular traffic models as suggested in TR 36.888~\cite{tr_36888}, whereas the SCUBA data arrivals are modeled as a busy Poisson traffic in all our simulations to evaluate the worst-case performance. To better investigate the performance of SCUBA in special scenarios, e.g., for SCUBA-SAM, we consider additional test-cases with lengthy ConA duration. Thus we fundamentally classify the simulation cases into short-data ($t_\data=250$ ms) and long-data ($t_\data=5$ s) cellular traffic conditions. The long $t_\data$ in cellular traffic is used to evaluate the performance gain of SCUBA-SAM, i.e., the latency gains obtained using SAM-D. The simulation settings applicable to both the cases are given in Table~\ref{table:simulation_settings1} whereas the case-specific simulation settings are defined in Table~\ref{table:simulation_settings2}. For the primary RAT, we choose the values of network parameters such as RRC connection set-up time ($t_\rrc$) and $\drxinat$ conforming with the set of values allowed by the 3GPP specification~\cite{ts_36331}. Based on ERP regulations in Table~\ref{table:regulations}, and considering $45\%$ power amplifier efficiency and $60$~mW power consumption in the support circuitry~\cite{ericsson_tdoc}, we set the UE power consumption for SCUBA transmission as $100$~mW. For the SCUBA traffic, we select those MCS and TBS values from TS~36.213~\cite[Table 7.1.7.2.1-1, Table 8.6.1-1]{ts_36213} that are compliant with regulations~\cite{rajendran2020}. For the analysis and simulations, we choose an MCS of 6, TBS of $256$~bits corresponding to a PRB size of 3, and cross-layer overhead of $19$~bytes, for the SCUBA traffic to comply with duty-cycle limits~\cite{rajendran2020}. \textcolor{black}{We provide a brief justification for the chosen MCS and PRB size in the appendix.} The selected MCS value corresponds to data modulated with quadriphase shift keying (QPSK) and coded with a rate of approximately 0.3. For power and latency simulations, we consider only a single pair of SRC and DST UEs and hence the results do not reflect the impact of collisions resulting from different UEs transmitting to the same DST. We show in Section~\ref{sec:results_collision} that this is a valid assumption for most MTC system architectures. We focus on the unlicensed bands of $865-868$~MHz and $902-928$~MHz for Europe and  United States, respectively, and therefore, a bandwidth of at least $3$~MHz is available for SCUBA. Since we choose LTE-M as the primary RAT for simulations, the signals are limited to 6 PRBs ($1.08$~MHz). Thus the considered band allows us to have a minimum of $2$ non-overlapping signal bandwidths. Hence for the collision evaluation, we further set $N_{\text{B}}=2$ to investigate the worst-case. When fewer PRBs are used and/or larger bandwidth is available for SCUBA resulting in a higher $N_{\text{B}}$, the collision rates are scaled down accordingly.


\begin{table}[t]
	\centering
	\caption{Common Simulation Settings~\cite{liberg2018mtcbattery, ericsson_tdoc, ericsson_tdoc1}} \label{table:simulation_settings1}
	\begin{tabular}{c|c||c|c}
		\hline
		{Parameter} & {Value} & Parameter & Value \\ \hline \hline 
		$\gamma^{-1}=\lambda_{\text{TXData}}^{-1}=\lambda_{\text{RXData}}^{-1}$ & 30 s & $P_\tx$ & 100 mW\\ \hline
		$P_\switch$ & 80 mW & $P_\rx$ & 80 mW\\ \hline
		$n_\slinat$ & 0 & $n_\sl$ & 8 \\ \hline
		$N_{\sam}$ & 150 ms & $n_{\sam}$ & 0.5 ms \\ \hline
		$N_{\samd}$ & 75 ms & $n_{\slpo}$ & 4 \\ \hline
		$N_\samu$ & 20 ms & $N_{\text{B}}$ & 2 \\ \hline 
		Mean~$\iat$ & 30 s &  $\iat$ & 5 min \\
		(Poisson, primary RAT) & & (periodic, primary RAT) \\ \hline 
		CDRX-ON duration & 20 ms & CDRX cycle & 640 ms \\ \hline
		$t_\rrc$ & 100 ms & IDRX cycle & 640 ms \\ \hline
		$\drxinat$ & 100 ms & RAI enabled & No \\ \hline
	\end{tabular}
\end{table}

\begin{table}[t]
	\centering
	\caption{Case-Specific Simulation Settings} \label{table:simulation_settings2}
	\begin{tabular}{c|c|c}
		\hline
		Case        &  Short $t_\data$ & Long $t_\data$ \\ \hline \hline 
		$t_\data$     & 250 ms  & 5 s   \\ \hline 
		$\datainat$ & 10 s   & 5 s    \\ \hline 
	\end{tabular}
\end{table}
\subsection{Simulation Results: Power and Latency}
\subsubsection{Short-Data Cellular Traffic}
\begin{figure}[t]
	\centering
	{\includegraphics[width=9cm]{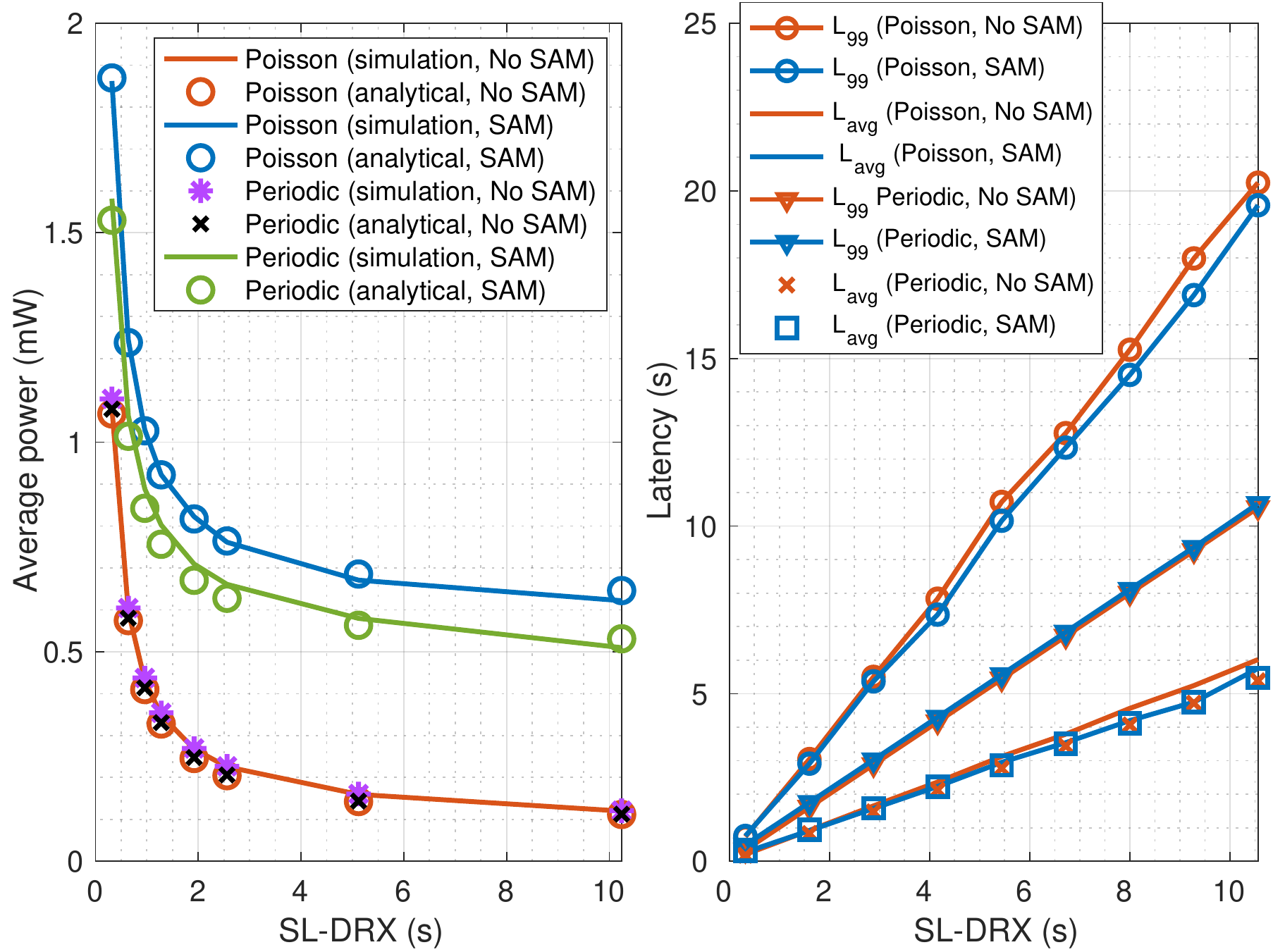}}\vspace{-.1in}
	\caption{Average power consumption vs SL-DRX (left), and latency vs SL-DRX for short data duration (right).}
	\label{fig:ShortPowerLatency}\vspace{0in}
\end{figure}

We present the SCUBA power consumption results in Fig.~\ref{fig:ShortPowerLatency} (left). Since the UE listens on SL-POs more frequently at lower SL-DRX cycle values, the power consumption significantly reduces with increase in SL-DRX cycle. The power consumed for SL data transmission and reception remains the same regardless of the SL-DRX cycle, and therefore has little impact on the power variation. For both Poisson and periodic traffic in primary RAT, native SCUBA has similar power consumption, since the underlying cellular traffic plays little role, except in blocking rare SCUBA traffic on occasion,  resulting in retransmission. The effect of such transmission failures and resultant retransmissions is negligible compared to the overall power consumption due to SL-PO listening. 

We show the latency incurred by the SCUBA traffic in Fig.~\ref{fig:ShortPowerLatency} (right), where $L_{99}$ and $L_\text{avg}$ correspond to 99th percentile and average latencies respectively. We observe that the latency increases nearly linearly with the SL-DRX cycle duration. Along with providing no  improvement in latency, using SAM results in higher power consumption due to SAM transmission and listening. In the presence of SAM, the power consumption in Poisson traffic model is high compared to the periodic case because of the higher number of SAM transmissions due to dominant ConA and CDRX modes. This clearly indicates that the use of SCUBA-SAM is not suitable when UEs operate with short data lengths in the primary RAT.
\subsubsection{Long-Data Cellular Traffic}
\begin{figure}[t]
	\centering
	{\includegraphics[width=9cm]{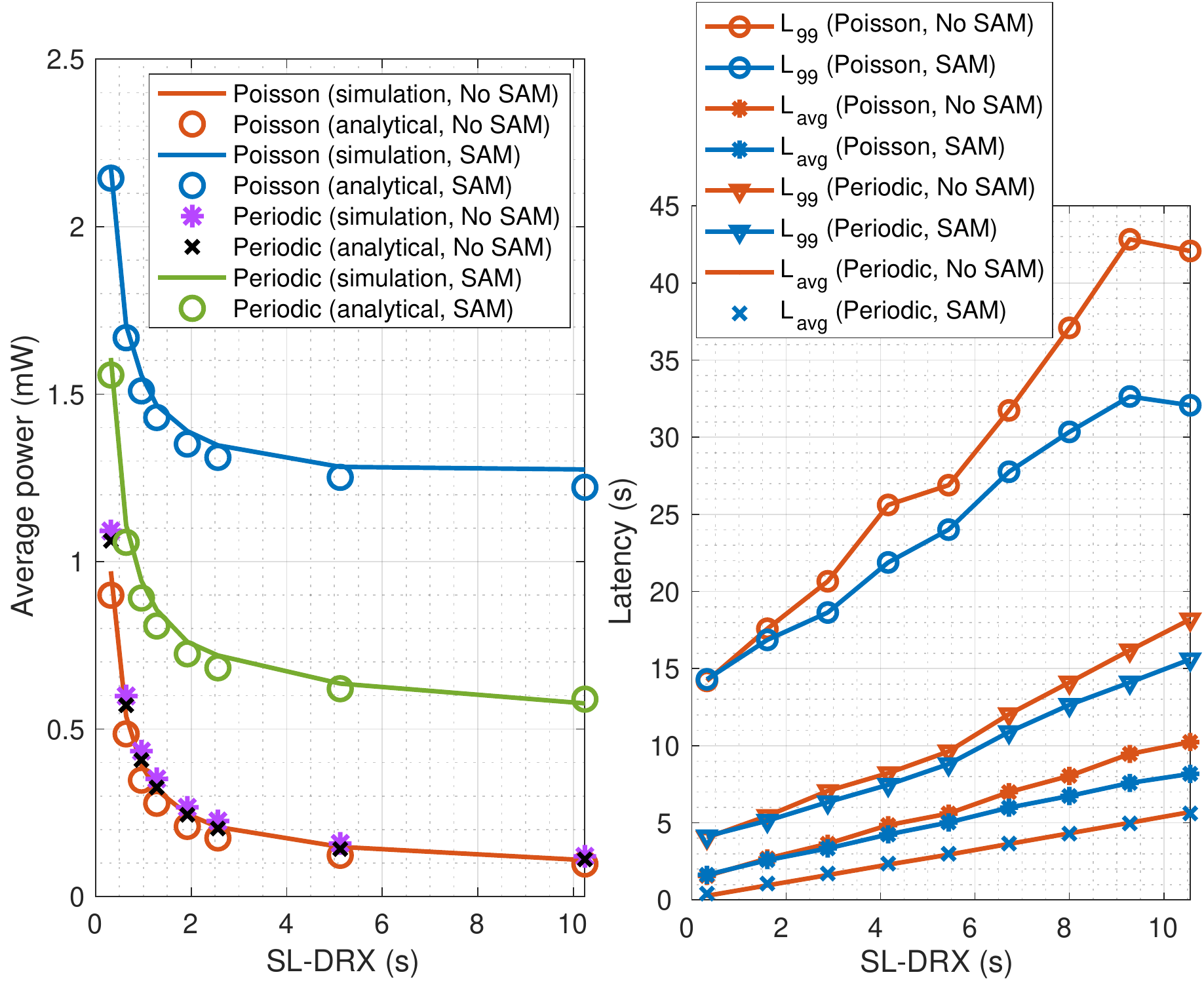}}\vspace{-.1in}
	\caption{Average power consumption vs SL-DRX (left), and latency vs SL-DRX for long data duration (right).}
	\label{fig:LongPowerLatency}\vspace{0in}
\end{figure}

The results of power consumption and network latency for long data traffic in the primary RAT are shown in Fig.~\ref{fig:LongPowerLatency}. Under the absence of SAM, SCUBA power remains the same as in the short-data case since there are no additional factors which contribute to the power consumption. However, SAM in long-data case results in much higher power consumption than in short-data case due to the higher percentage of ConA and CDRX modes causing more number of SAM transmissions. This increased power consumption due to SAM is also associated with an improvement in the achieved latency. 
The achieved latency gain increases with SL-DRX cycle duration, and is higher for SCUBA multiplexed with Poisson cellular traffic as compared to periodic cellular transmission. The result shows that a gain of more than $23\%$ is observed in 99th percentile latency with the use of SAM when the SL-DRX cycle is $10.56$~s. 
The results clearly show that SCUBA-SAM is particularly useful when the UEs are busy with cellular traffic causing ConA and CDRX modes to occupy a majority of the time. In busy cellular traffic, the ConA and CDRX-ON modes overlap with most of the fixed SL-POs, which introduces higher latency. This latency can be reduced with the use of SAM by assisting the SRC to transmit on an earlier dynamic SL-PO without waiting for the next fixed SL-PO.

To avoid repetition, we do not show the simulation results for the case where SAM is transmitted in cellular IDRX mode when RAI is enabled. We observed similar latency gains for SAM mode in the long-data case when RAI is enabled because of SAM-D in IDRX allowing for an early SL transmission.

\begin{table}[]
	\centering
	\caption{Latency simulation results for LLM} \label{table:LLMLatency}
	\begin{tabular}{l|l|l|l|l}
		\hline
		\multicolumn{1}{c|}{\multirow{2}{*}{Cellular traffic}} & \multicolumn{2}{c|}{Short data} & \multicolumn{2}{c}{Long data} \\ \cline{2-5}
		\multicolumn{1}{c|}{}    & 99 percentile      & Average     & 99 percentile      & Average     \\ \hline \hline 
		Poisson              & 340.8 ms         & 28.2 ms        & 14.11 s          & 1.392 s        \\ \hline 
		Periodic             & 38.0 ms           & 20.6 ms        & 3.717 s          & 108.9 ms         \\ \hline 
	\end{tabular}
\end{table}

\begin{table}[t]
	\centering
	\caption{Average power consumption results for LLM} \label{table:LLMPower}
	\begin{tabular}{l|l|l|l|l}
		\hline
		\multicolumn{1}{c|}{\multirow{2}{*}{Cellular traffic}} & \multicolumn{2}{c|}{Short data} & \multicolumn{2}{c}{Long data} \\ \cline{2-5}
		\multicolumn{1}{c|}{}    & Simulation      & Analysis     & Simulation     & Analysis     \\ \hline \hline 
		Poisson              & 78.2 mW         & 78.6 mW        & 67 mW          & 67 mW        \\ \hline 
		Periodic             & 79.8 mW           & 80.1 mW        & 78.6 mW          & 78.9 mW         \\ \hline 
	\end{tabular}
\end{table}

Since the SL-DRX cycle is $0$ for SCUBA-LLM, we tabulate the latency and power values in Table~\ref{table:LLMLatency} and Table~\ref{table:LLMPower} separately, respectively. LLM latency is significantly smaller for SCUBA multiplexed with the longer IAT periodic data transmission than a higher rate of arrival Poisson cellular traffic, because the latter hinders SL listening more with frequent data arrivals. We observe that the SCUBA-LLM power consumption is higher in case of short cellular data than in the longer data traffic case. The power difference is also significant for Poisson cellular traffic, whereas it is marginal with periodic arrivals because the former allows little SL listening resulting from more frequent long data arrivals in the primary RAT. 
Overall, the results show that SCUBA-LLM allows nearly instant SL communication at the expense of high power consumption compared to the version of SCUBA with non-zero SL-DRX cycles.
\textcolor{black}{\subsubsection{Comparison with Other D2D Solutions}
To put the performance of SCUBA in perspective, we consider the power consumption values of other D2D technologies reported in the literature~\cite{bluetoothpower}. For comparison with SCUBA, we focus on Bluetooth low energy (BLE) and ZigBee, since they consume lower power than the other state-of-the-art commercialized D2D solutions~\cite{bluetoothpower}. Note that the transmitter power we have chosen for SCUBA is the maximum transmit power allowed in the $865-868$~MHz band. The power consumption in SCUBA will however be different when transmitting in a different regulated unlicensed band.}

\begin{figure}[t]
\textcolor{black}{
	\centering
	{\includegraphics[width=9cm]{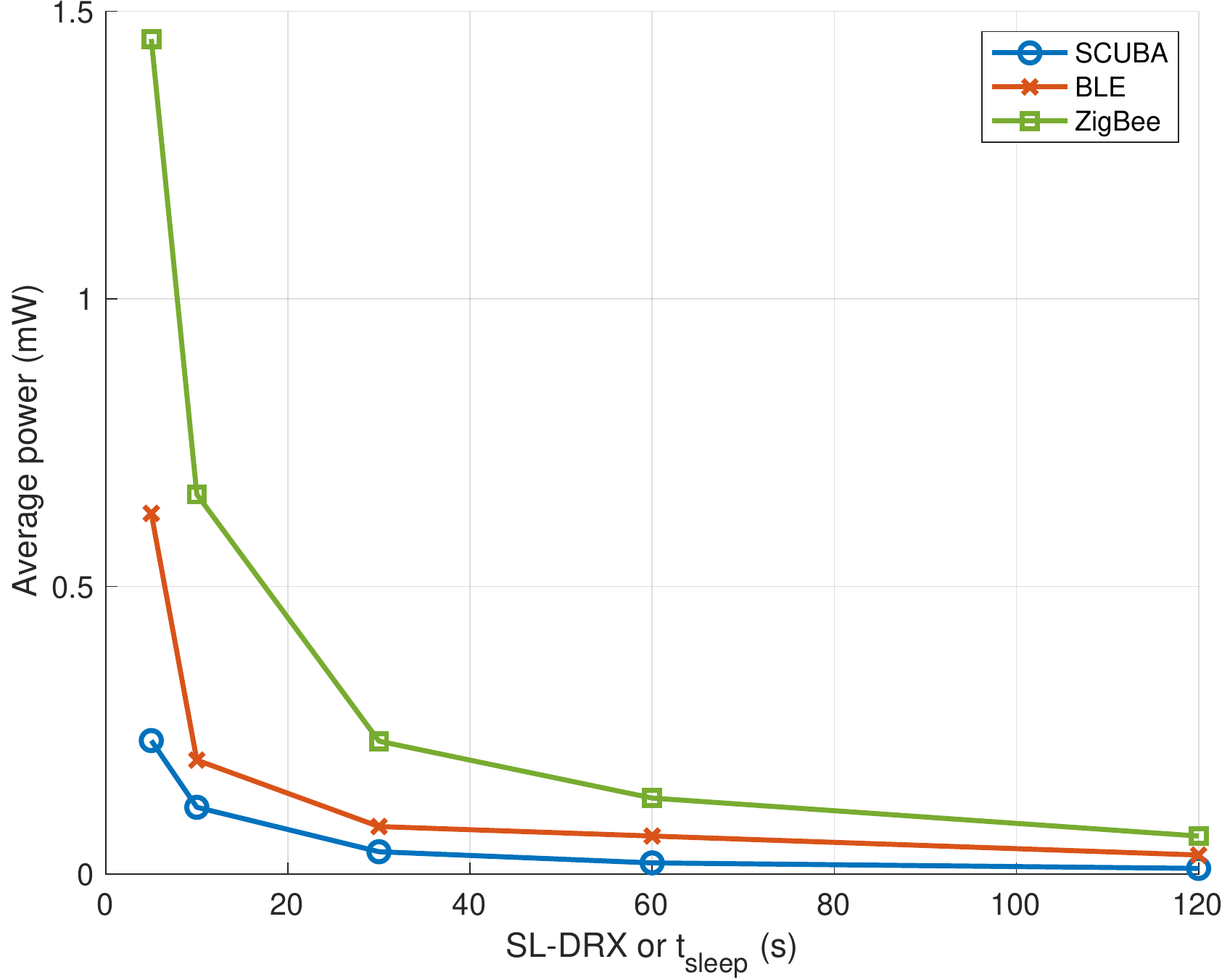}}\vspace{-.1in}
	\caption{Average power consumption vs SL-DRX or $\text{t}_{\text{sleep}}$ for native SCUBA, BLE, and ZigBee~\cite{bluetoothpower}.}
	\label{fig:PowerComparison}\vspace{0in}}
\end{figure}

\textcolor{black}{Prior to comparing SCUBA with other D2D solutions, we emphasize two major differences in the nature of the protocols. First, transmit power of signals in BLE and ZigBee are regulated to be fixed between $1-100$~mW~\cite{bluetooth_spec}, which severely limits the achievable coverage when compared to that of SCUBA. Furthermore, unlike SCUBA, BLE and ZigBee are incompatible to function as a TDM solution in low-cost LTE-M HD-FDD UEs using a single radio architecture. Nevertheless, the power consumption of BLE and ZigBee devices are shown in Fig.~\ref{fig:PowerComparison}. We present the average power consumption comparison between BLE, ZigBee, and native SCUBA over different sleep cycles. The power consumption values of BLE and ZigBee devices are from~\cite{bluetoothpower}, and correspond to periodic $8$~byte data transmissions at $1$~mW between sleep cycles of $t_{\text{sleep}}$ duration. On the other hand, the SCUBA power consumption results correspond to an LTE-M device transmitting an $8$~byte data at a transmit power of $25$~mW. The power consumption in SCUBA is evidently lower than BLE and ZigBee in spite of latter solutions having lower transmit power. Since SCUBA operates as a secondary RAT utilizing the synchronization acquired through primary RAT, and transmits on pre-defined SL-POs, it does not require power-expensive advertising procedures as in BLE and Zigbee~\cite{bluetoothpower}. Considering the higher transmission power, SCUBA also has the potential to provide much wider coverage than BLE and ZigBee .} 

\textcolor{black}{Additionally, SCUBA also facilitates LLM for latency-critical applications, where it provides the earliest possible transmission opportunity while coexisting with a primary RAT. For a transmission of $376$~bits, SCUBA LLM average latency for cellular periodic short-data case is $9.97$~ms against the combined discovery-connection mode average latency of $11$~ms in BLE~\cite{bluetoothlatency2, bluetoothlatency1}. Furthermore, unlike BLE which operates only in $2.4$~GHz band, SCUBA can function in any unlicensed frequencies of choice, by following the regulatory requirements set for each band~\cite{rajendran2020}. In summary, different from other existing D2D standards, SCUBA can coexist with a cellular RAT in a TDM manner with considerably lower power consumption, improved coverage expansion, and comparable latency performance.}
\subsubsection{Battery life}
Since we have chosen LTE-M as the primary RAT for our SCUBA simulations and analysis, we use the energy consumption analysis of MTC devices provided in the 3GPP technical report~\cite{tr_45820} to estimate the impact of SCUBA on the battery life of a UE. We consider a UE having an ideal battery source of capacity $E_{\text{b}} = 5~\text{Wh}$ with no power leakage~\cite{tr_45820}. The energy consumption per day by SCUBA and LTE-M operating together can be computed as
\begin{equation}
E_{\text{day}}=E_{\text{SCUBA}}+E_{\text{LTE-M}}~\text{Wh},
\end{equation}
where $E_{\text{SCUBA}}$, $E_{\text{LTE-M}}$ are per-day energy consumption by SCUBA, LTE-M respectively. Then, the UE battery life can be computed as, $D=\frac{E_{\text{b}}}{E_{\text{day}}}$ days. 

For SCUBA, first we consider Poisson traffic with the same values of parameters as in Table~\ref{table:simulation_settings1}. Since LLM and SAM are intended for delay sensitive applications tailored for UEs mostly connected with alternating current (AC) power source or devices without battery-life constraints, we analyze the battery life impact of SCUBA in its native mode only. For LTE-M, we consider the battery-life analysis as given in the literature~\cite{Xu2018battery, liberg2018mtcbattery}. For LTE-M standalone operation, if a device with battery capacity of $5$ Wh transmits packets worth $200$ bytes in a cellular link at $164$~dB maximum coupling loss (MCL) at a frequency of $2$ hours, the battery lasts for $328.5$ days~\cite{liberg2018mtcbattery}. SCUBA power consumption values are independent of the underlying cellular traffic. Based on the average SCUBA power consumption value for SL-DRX cycle of $10.24$~s obtained from simulation results (Fig.~\ref{fig:ShortPowerLatency}), we estimate the UE battery life for LTE-M-SCUBA coexisting scenario. We find that the battery life reduces to $279$ days when SCUBA coexists together with LTE-M. Note that these numbers are obtained for a fairly busy SCUBA traffic of Poisson packet arrivals with mean IAT of $30$~s, and are therefore indicative of the worst-case scenario. 

For a more realistic practical indication, we also investigate the battery life when SCUBA traffic is as infrequent as in LTE-M, i.e., IAT of $2$~hours. The battery life is found to be $328.3$ days for SCUBA-LTE-M coexistence as compared to $328.5$~days in the absence of SCUBA. This shows the effectiveness of the low-power design of SCUBA, where it consumes power only for SL transmissions and receptions, and reuses the synchronization achieved in the cellular link. This energy consumption performance highlights the appeal of SCUBA to be ideally suited as an integrated low-cost and low-power solution for SL-U in low-complexity devices, particularly LTE-M Cat-M1 UEs used for MTC and IoT applications.
\subsection{Simulation Results: Collision}\label{sec:results_collision}
\begin{figure}[t]
	\centering
	{\includegraphics[width=8.5cm]{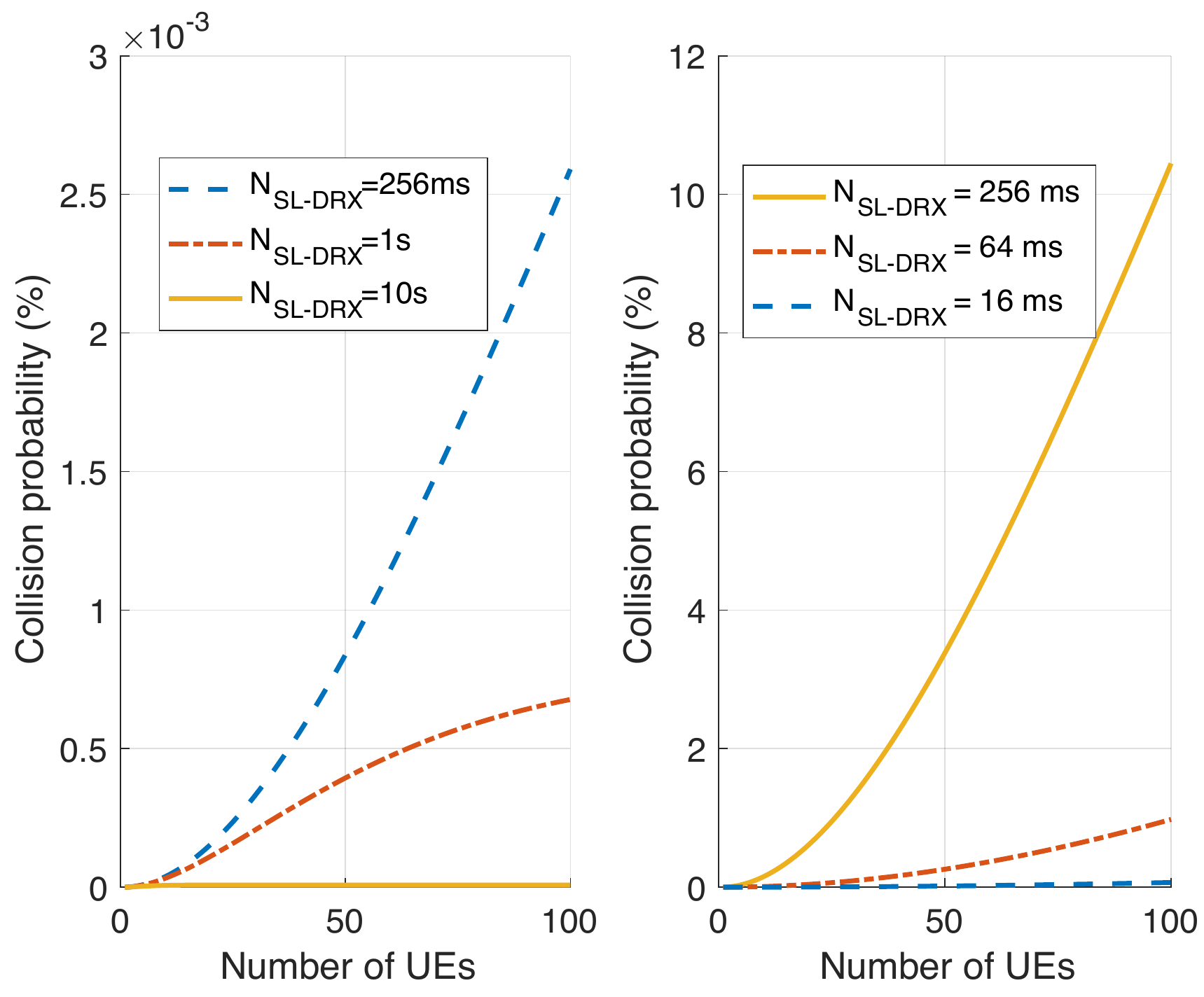}}
	\caption{SCUBA data collision probability when UEs communicate with each other randomly (left), and when all SRC UEs always report to a central DST (right).}
	\label{fig:slDataColl} 
\end{figure}
For power and latency simulations discussed in the previous sections, we did not consider the collisions resulting from different UEs transmitting at the same SL-PO of  a DST. If collisions occur in SCUBA transmission, the power consumption and latency will increase accordingly. In this section, we present the collision results obtained using the analysis in Section~\ref{section:packetcollisions}. We first show the results of SCUBA data collisions in Fig.~\ref{fig:slDataColl}. As expected, we observe an increase in the rate of collision with the number of participating UEs. For the case of devices communicating with each other at random, seen in Fig.~\ref{fig:slDataColl} (left), we notice a higher probability of collision with lower SL-DRX cycle as it increases the possibility of SL-PO overlap among DST UEs. The absolute value of collision rates under all these conditions are in the order of $10^{-5}$, indicating a nearly collision-free SCUBA communication across different operating conditions. However, when all source UEs are reporting to a common central DST, the collision rates increase with higher SL-DRX cycle, as seen in Fig.~\ref{fig:slDataColl} (right), since a longer interval between SL-POs allows for more SRC UEs to have SL data to transmit. Therefore, DST UEs that are expected to receive data from several/all network nodes should use a lower SL-DRX cycle to avoid SCUBA packet collisions. Note that these collision results are independent of short- or long-data in the primary RAT, since we evaluate the worst-case collision with $p(C)=1$. \textcolor{black}{Given that the SCUBA packet collision rates are negligible for both categories of network architectures, we do not compare these values against the probabilities of collision obtained with other D2D solutions, since it does not provide any further meaningful insights.}

\begin{figure}[t]
	\centering
	{\includegraphics[width=7cm]{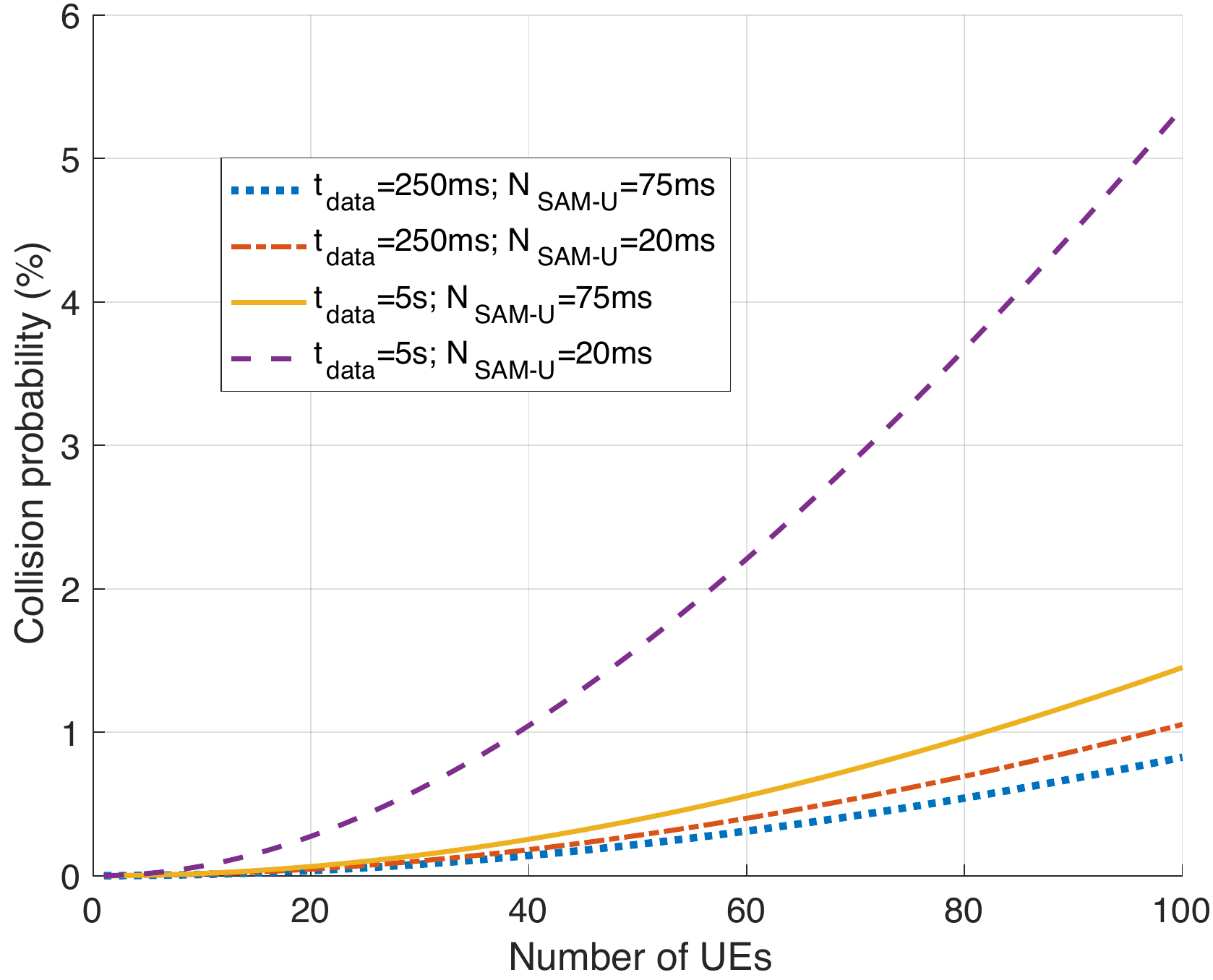}}
	\caption{SAM collision probability for SCUBA-SAM mode.}
	\label{fig:sam_coll} 
\end{figure}
Similarly, we present the results of SAM collisions for the case of SCUBA-SAM in Fig.~\ref{fig:sam_coll}. We observe that type of traffic in the primary RAT significantly impacts the collision rates, with long-data at the cellular link resulting in higher SAM collisions and short-data causing fewer collisions. This behavior is a result of UE spending a higher amount of time in ConA mode, during which it regularly transmits SAM-U. These results indicate that the minor gains in network latency (less than $50$~ms) achievable with use a SAM transmission interval of $20$~SFs incurs a high cost of collision of up to a $3\times$ increase in collision rates, especially for the case of long-data. This suggests that using a higher SAM transmission interval of, e.g., $N_\samu = 75$~ms is suitable. It also ensures that at least $2$ SAMs are transmitted within a SAM period of $150$~SFs, which nearly guarantees that a listening UE does not miss a SAM due to collision.

\section{Discussion}\label{section:discussion}
In this section, we reflect on SCUBA by discussing its salient features and identifying potential future work required for a practical realization.     


\subsection{SCUBA Customization}
SCUBA is a customizable protocol that can be further adapted based on the application scenario. For example, native SCUBA in its current form works on UEs that are in one of the DRX modes for cellular communication. However, if their ConA transmission is deterministic in nature, i.e., if the SRC is aware a-priori of idle times in ConA (e.g., switch SF) of the DST, SCUBA can also further be modified to transmit SL data to a UE in ConA mode, and subsequently expect an ACK back in the next idle time of the DST. 

\subsection{Other Layer Specifications}
This paper focuses on the MAC layer specifications of SCUBA. Its PHY characteristics are largely driven by the regulations governing the access of the unlicensed spectrum used. For example,~\cite{rajendran2020} lists the PHY specifications for the use of $865-868$~MHz band in Europe and $902-928$~MHz band in the United States. For upper layer specifications, SCUBA borrows them from the underlying primary RAT used in the UE. For LTE-MTC devices, SCUBA uses the radio link control, radio resource control, packet data convergence control, and the non-access stratum protocols from LTE-M.

\subsection{Standardization of SCUBA}
SCUBA is an ideal candidate to be integrated into the fold of MulteFire specifications, which primarily provides solutions for operating cellular technologies in shared and unlicensed spectrum~\cite{MF1}. In addition, specific features of SCUBA find applicability in enhancing the performance of 3GPP LTE D2D and 5G new radio SL (NR SL). For example, 3GPP plans to standardize ACK/negative-ACK (NACK)-based HARQ feedback scheme to ensure reliability for NR vehicle-to-everything (V2X) protocol which is built upon NR SL~\cite{tr_37985}. Similarly, the notion of SL DRX has been incorporated currently as a work item by 3GPP to be introduced in NR SL~\cite{NRSL_WID}. The targets of the work item include designing collision avoiding and device-aware SL-POs, similar to the idea we introduce for SCUBA in Section~\ref{section:sldrx}.

\section{Conclusion}\label{section:conclusion}
SCUBA enables direct communication between cellular devices in the unlicensed frequency bands. It offers the unique benefit over the state-of-the-art unlicensed D2D RATs that it coexists with the underlying legacy cellular protocol while \textit{reusing the existing hardware}. We provide the PHY/MAC layer specifications of SCUBA, including the SL-DRX technique that provides an application controlled flexibility between latency and power consumption. SCUBA also includes optional features of a low latency mode for near instantaneous SL transmission and a SCUBA-SAM mode for achieving reasonable SL latency targets under busy cellular traffic conditions. The collision rates, power consumption, and latency analyses and simulation results prove that SCUBA is an appealing low-cost and low-power solution for SL communication on unlicensed bands.


\appendix
\textcolor{black}{The choice of MCS and PRB size impacts the SCUBA network latency. To evaluate the worst case performance, we consider the lowest possible MCS and the fewest PRBs, which result in the slowest possible transmission. However, the choice of MCS and the number of PRBs is further restricted by the following regulations governing the use of unlicensed bands.} 

\textcolor{black}{The North American regulations impose that the $6$~dB bandwidth of all digitally modulated systems must be greater than $500$~kHz without any restriction on the maximum usable bandwidth~\cite{CFR_FCC}. With LTE-M as the underlying primary RAT whose parameters and specifications are also reused for SCUBA, a minimum of three PRBs are required to ensure that the $6$~dB bandwidth is greater than $500$~kHz.}

\textcolor{black}{For the choice of MCS, we consider the duty cycle limitations imposed by European unlicensed band usage regulations. Our prior analysis has established that the minimum MCS that satisfies the duty cycle regulations is MCS~$\geq 4$~\cite{rajendran2020}. However, this analysis considers that the transmissions are error-free. For practical communication conditions with non-negligible block errors, erroneous transport blocks are typically retransmitted. The number of transport blocks retransmitted is dependent on the operating signal-to-noise ratio conditions, the chosen target block error rate, and the retransmission scheme used, e.g., whether Layer-1 HARQ is used or RLC retransmissions are used. To accommodate for several possible retransmissions, we choose a retransmission-agnostic conservative choice of MCS~$=6$ for our evaluations. With the chosen MCS and number of PRBs, we extract the TBS from ~\cite[Table 7.1.7.2.1-1, Table 8.6.1-1]{ts_36213}.}

\bibliographystyle{IEEEtran}		
\bibliography{References}{}

\end{document}